\documentclass[preprint,superscriptaddress,nofootinbib,preprintnumbers,amsmath,amssymb,11pt]{revtex4-1}
\usepackage[utf8]{inputenc}
\usepackage{amsfonts}
\usepackage{mathrsfs}
\usepackage{physics}
\usepackage{footnote}
\usepackage{scrextend}
\usepackage{leftidx}
\usepackage{soul}

\usepackage[dvipsnames]{xcolor}
\definecolor{red}{rgb}{0.9, 0,0}
\definecolor{cerulean}{rgb}{0., 0.42,0.9}

\usepackage{epsfig}
\usepackage{graphicx}  
\usepackage{url}
\usepackage{color}
\usepackage{hyperref}
\hypersetup{
     colorlinks   = true,
     citecolor    = red,
	 linkcolor=blue
}
\usepackage{float}

\newcommand{\bfa}{{\bf a}}
\newcommand{\bfe}{{\bf e}}
\newcommand{\bfx}{{\bf x}}
\newcommand{\bfq}{{\bf q}}

\newcommand{\bfr}{{\bf r}}
\newcommand{\bfl}{\ensuremath{\boldsymbol\ell}}
\newcommand{\bfk}{{\bf k}}
\newcommand{\bfG}{{\bf G}}
\newcommand{\bfv}{{\bf v}}
\newcommand{\bfu}{{\bf u}}

\renewcommand{\bra}{\langle}
\renewcommand{\ket}{\rangle}

\definecolor{red}{rgb}{0.9, 0,0}
\definecolor{cerulean}{rgb}{0., 0.62,0.9}
\definecolor{navy}{rgb}{0.05, 0.05,0.8}

\graphicspath{{./figures/}}

\begin{document}

\title{Multiphonon excitations from dark matter scattering in crystals}
\author{Brian Campbell-Deem}
\affiliation{ Department of Physics, University of California, San Diego, CA 92093, USA }
\author{Peter Cox}
\affiliation{School of Physics, The University of Melbourne, Victoria 3010, Australia}
\affiliation{Kavli Institute for the Physics and Mathematics of the Universe (WPI), UTIAS, The University of Tokyo, Kashiwa, Chiba 277-8583, Japan}
\author{Simon Knapen}
\affiliation{School of Natural Sciences, Institute for Advanced Study, Princeton, NJ 08540, U.S.A.}
\author{Tongyan Lin}
\affiliation{ Department of Physics, University of California, San Diego, CA 92093, USA }
\author{Tom Melia}
\affiliation{Kavli Institute for the Physics and Mathematics of the Universe (WPI), UTIAS, The University of Tokyo, Kashiwa, Chiba 277-8583, Japan}
\date{\today}

\begin{abstract}
\noindent 
For direct detection of sub-MeV dark matter, a promising strategy is to search for individual phonon excitations in a crystal.
We perform an analytic calculation of the rate for light dark matter \mbox{(keV $<m_{DM}<$ MeV)} to produce two acoustic phonons through scattering in cubic crystals such as GaAs, Ge, Si and diamond. The multiphonon rate is always smaller than the rate to produce a single optical phonon, whenever the latter is kinematically accessible.  In Si and diamond there is a dark matter mass range for which multiphonon production can be the most promising process, depending on the experimental threshold.
\end{abstract}

\maketitle

\tableofcontents
\clearpage
\section{Introduction}

\subsection{Motivation}

The quest to directly detect dark matter (DM) in a laboratory experiment has, in recent years, significantly diversified~\cite{Battaglieri:2017aum}; both theoretical and experimental developments have driven the search beyond the WIMP paradigm. A steady decrease in energy thresholds has enabled sensitivity to particle-like dark matter with a mass well below that of a typical WIMP, with masses as low as an MeV currently being probed. Next generation detectors aim to push down to the lower limit of particle-like dark matter, probing the keV-MeV mass range. The energy scales for excitations created in current and proposed detectors coincide with the energy scales typical of many-body excitations in condensed matter or atomic systems. For DM heavier than $\gtrsim 1$ MeV, electronic excitations in atoms or semiconductors with energy gaps in the $\sim$ eV range are well suited, if the DM couples to electrons \cite{Agnese:2018col,Abramoff:2019dfb,Aguilar-Arevalo:2019wdi,Agnes:2018oej,Essig:2015cda,Essig:2017kqs,Hochberg:2016ntt,Griffin:2019mvc}. For light DM with nucleon couplings on the other hand, one can utilize chemical bond breaking \cite{Essig:2016crl}, nuclear de-excitations \cite{Pospelov:2019vuf}, crystal defects \cite{Kadribasic:2017obi,Budnik:2017sbu} or soft nuclear recoils, where the latter in particular require very low thresholds \cite{Angloher:2015ewa,Aguilar-Arevalo:2016ndq,Arnaud:2017bjh,Agnese:2016cpb,Guo:2013dt,Hertel:2018aal,Kurinsky:2019pgb}. For DM lighter than 1 MeV, vibrational modes in crystals \cite{Knapen:2017ekk,Griffin:2018bjn,Trickle:2019nya,Griffin:2019mvc}, molecular systems \cite{Essig:2019kfe,Arvanitaki:2017nhi} or superfluid helium \cite{Schutz:2016tid,Knapen:2016cue,Acanfora:2019con,Caputo:2019cyg} naturally have energy spectra in the required 1 - 100 meV range. Possible alternative detection strategies in this mass range are electronic systems with ultra-low bandgaps \cite{Hochberg:2015pha,Hochberg:2015fth,Hochberg:2017wce,Coskuner:2019odd,Geilhufe:2019ndy}, magnon excitations \cite{Trickle:2019ovy} and avalanche gains in molecular magnets \cite{Bunting:2017net}.

On the theoretical side, it is necessary to understand dark matter interactions with vibrational modes rather than with individual nuclei, so as to reliably estimate sensitivity and, in the event of any signal, extract dark matter properties. The reason is that for DM lighter than $\sim$ MeV, its de Broglie wavelength exceeds the interparticle spacing in typical materials, and it becomes necessary to transition to a different effective theory by integrating out the nuclei and electron clouds. One can therefore expect new and interesting features in these interactions, as they are subject to different kinematics and symmetry principles than those which govern the interactions in conventional dark matter experiments.  

In this work, we focus on theoretical calculations for DM to excite  vibrational modes (phonons) in crystals.
In a crystal with a non-trivial primitive-cell, phonons can be characterised as either acoustic or optical. The acoustic phonons are the Nambu-Goldstone modes associated with the breaking of translation symmetry by the crystal lattice; they must, at low energies, obey a linear dispersion relation. This feature in particular poses an experimental challenge for the detection of DM with mass below $\sim$100 keV: the momentum transfer in this regime is comparatively low, and the linear dispersion relation of the acoustic branch with typical slope $\sim 10^{-5}$ implies a very low energy transfer ($\sim$ meV). The optical phonons, on the other hand, are gapped and have $\gtrsim 10$ meV of energy at arbitrarily low momentum transfer. This makes them experimentally much more attractive.

There are also important theoretical differences between acoustic and optical phonons, as their couplings to DM depend very strongly on the DM model~\cite{Knapen:2017ekk,Griffin:2018bjn,Cox:2019cod}. Concretely, if the DM has a coupling proportional to electric charge, there is a dipole interaction with the optical branches while the coupling to the acoustic branches is strongly suppressed. If instead the DM has a coupling proportional to the atomic mass, the coupling to the optical branches is strongly suppressed. At the same time, detecting single acoustic phonons is expected to be extremely challenging experimentally. This motivates the study of processes where \emph{multiple} phonons are produced, which can have larger energy transfer. This was studied already in the context of superfluid helium \cite{Schutz:2016tid,Knapen:2016cue,Acanfora:2019con,Caputo:2019cyg}, where the sound speed is particularly low and multiphonons were found to extend the reach for sub-MeV DM. 

The purpose of this paper is to compute multiphonon processes for cubic crystals such as Ge, Si, GaAs and diamond in the isotropic approximation. Such materials are either already being used or considered for direct detection experiments, and it was found previously that the isotropic limit matches the numerical result well for single phonon excitations in GaAs~\cite{Griffin:2018bjn}. More complicated, strongly anisotropic crystals, such as sapphire, are left for future work. We focus on DM that couples proportional to atomic mass number of the target nuclei, as it is in this scenario where multiphonon corrections are the most important. We focus on two acoustic phonons in the final state, for which there is a well-known effective theory, and briefly comment on multiphonon excitations with optical phonons.

\subsection{Summary of results}

\begin{figure}\centering
 \includegraphics[width=0.3\textwidth]{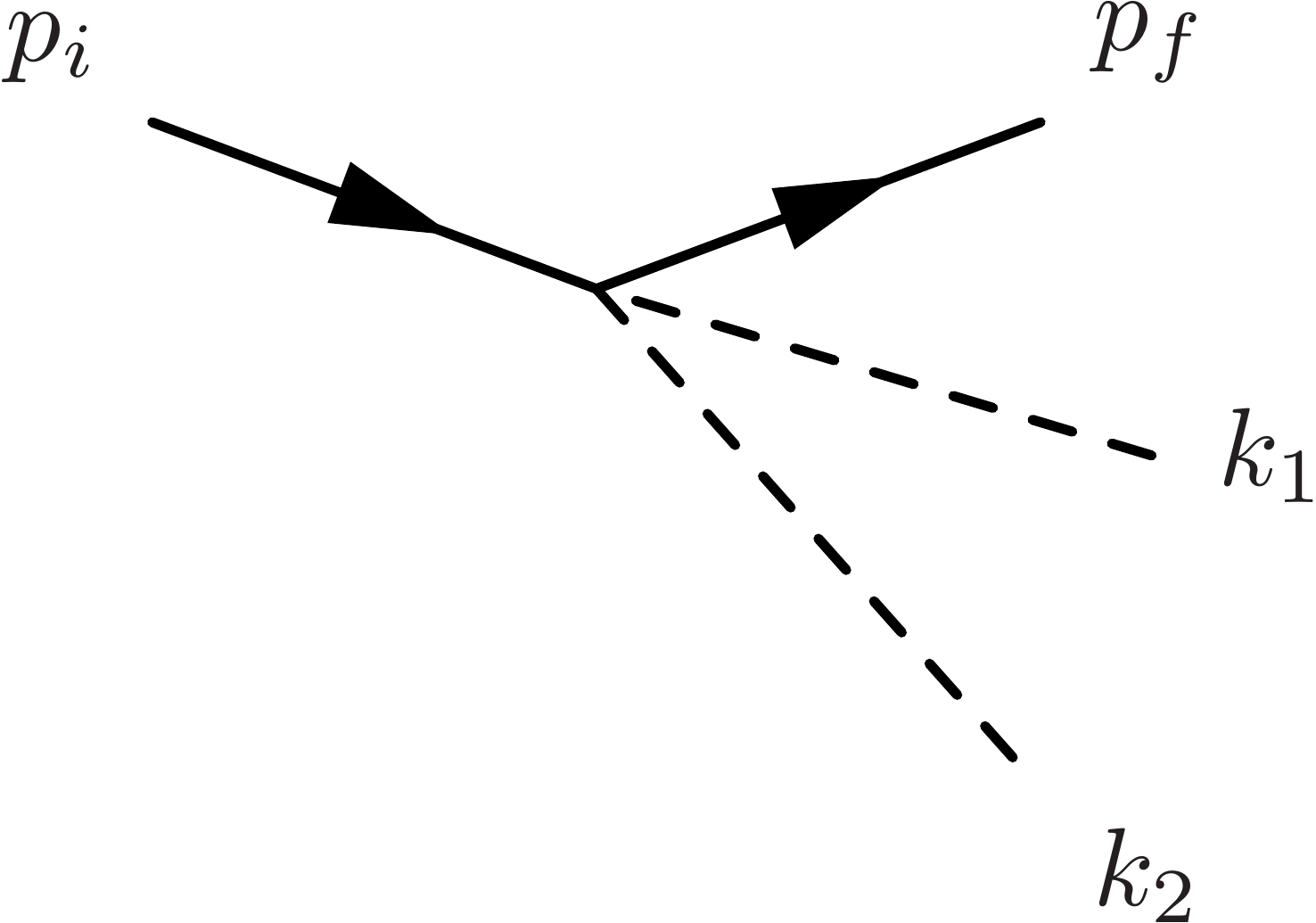}\hspace{2cm}
 \includegraphics[width=0.3\textwidth]{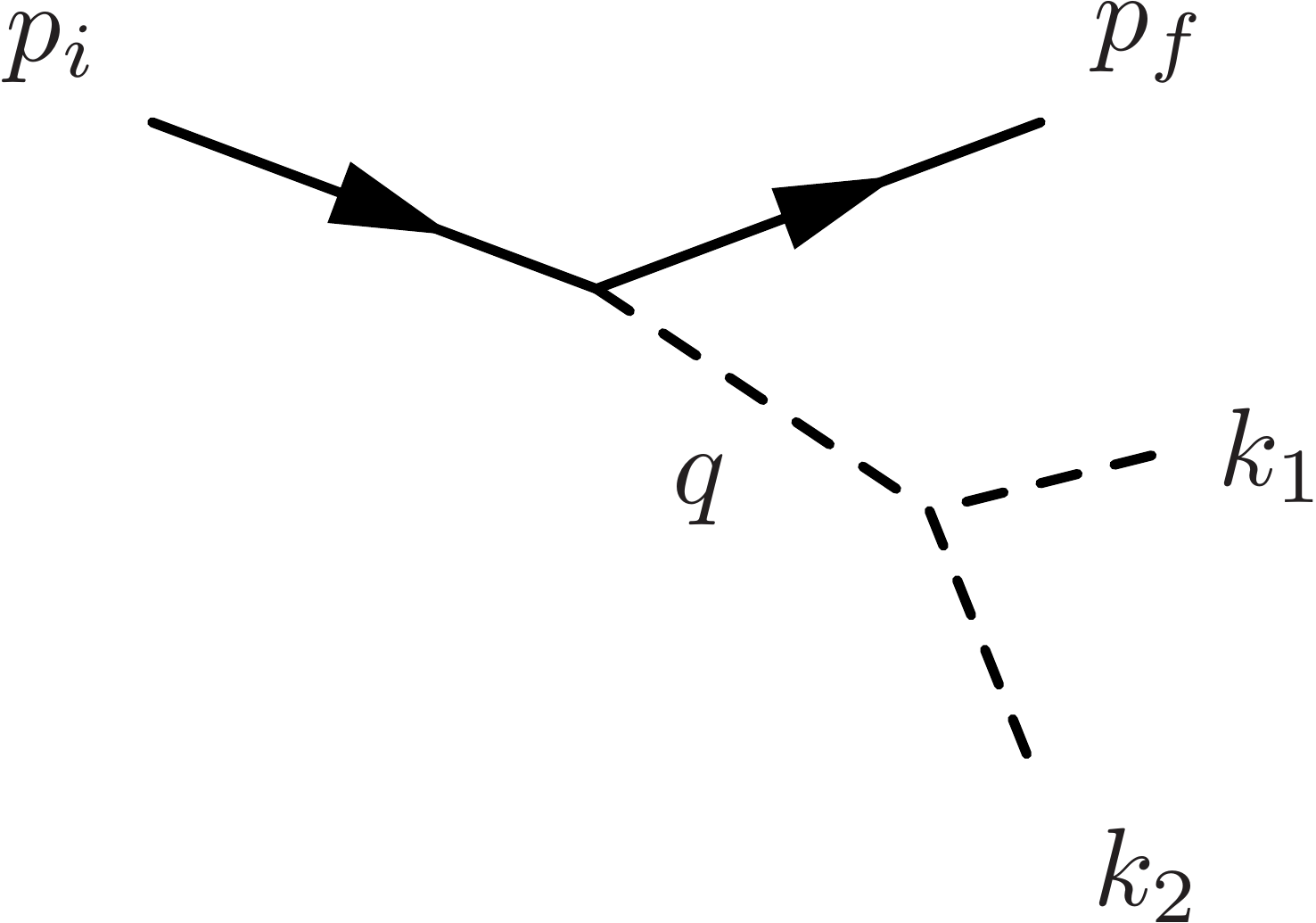}
 \caption{Diagrams representing the contact (left) and anharmonic (right) contributions to the DM scattering rate into two phonons (dashed lines). \label{fig:diagrams}}
\end{figure}

The main object we are computing is the structure factor $S(q,\omega)$, which parametrizes the scattering rate of an external probe to the crystal for a momentum transfer $q$ and energy transfer $\omega$ (see Sec.~\ref{sec:structure-factor} for a precise definition). There are two distinct contributions to $S(q,\omega)$ from the production of two phonons, represented by the diagrams in Fig.~\ref{fig:diagrams}. The left-hand diagram relies on a contact interaction between the DM and two phonons, which originates from the matching between the low energy effective phonon theory and UV theory of nuclei and electrons. There are analogous operators with three, four or more phonons, for which each additional phonon comes with a factor of $q/\sqrt{m_N \omega}$, with $m_N$ the nucleus mass. For $m_{DM}<$ MeV, $q/\sqrt{m_N \omega}$ is a good expansion parameter, rendering the $\geq 3$ phonon contributions negligible. For higher DM masses, the breakdown of this expansion signals the transition to the regular nuclear recoil regime. A resummation procedure is needed in this transition regime, which we do not attempt in this paper. 
 
The right-hand diagram in Fig.~\ref{fig:diagrams} instead occurs via an off-shell phonon and phonon self-interactions, which arise in part from the anharmonicity of the crystal potential. While this diagram only relies on the DM coupling to a single phonon, and therefore appears to be lower order in $q/\sqrt{m_N \omega}$, there is an additional suppression in $q$ from the insertion of the phonon self-interaction. 
We will see in Sec.~\ref{sec:istropic} that in the low-momentum regime, the self-interactions of acoustic phonons are governed by multiple dimensionful parameters that are related to the elastic constants of the crystal. The sense in which the self-interactions are ``small'' can be most easily seen from the fact that the typical width of the longitudinal acoustic phonon, $\Gamma$, is very small compared to its energy,\footnote{Note this is different from superfluid He, where the phonon-roton self-interactions are much larger, but where the phonon decay is kinematically forbidden for part of the dispersion curve.} in other words $\Gamma/\omega\ll 1$.

\begin{table}[b]
\begin{ruledtabular}
\begin{tabular}{lccc}
channel & low-$q$ scaling & typical threshold needed & Ref.\\\hline\hline
single acoustic phonon& $q$ & 1 meV & \cite{Knapen:2017ekk,Griffin:2018bjn,Trickle:2019nya}\\
single optical phonon& $q^4$ & 25 meV & \cite{Knapen:2017ekk,Griffin:2018bjn,Trickle:2019nya,Cox:2019cod}\\
multi-phonon (contact)&  $q^4$ &5-10 meV&this work\\
multi-phonon (anharmonic) & $\# q^{4}$ &5-10 meV&this work\\
multi-phonon (helium) &$\#q^{4}$ & 1 meV\footnote{A superfluid He detector benefits from a natural, exothermic evaporation-absorption process, such that the effective threshold of the sensor itself may be $\sim$10 meV \cite{Guo:2013dt}.}  &\cite{Schutz:2016tid,Knapen:2016cue,Acanfora:2019con,Caputo:2019cyg} \\
\end{tabular}
\end{ruledtabular}
\caption{Leading scaling of the structure factor $S(q,\omega)$ in the low $q$ (low $m_{DM}$) limit for different channels, and required approximate thresholds to observe them. It is assumed that the DM couples proportional to the mass of the atoms. The $\#$ indicates that this channel vanishes in the limit where the (material dependent) phonon self-couplings are taken to zero. \label{tab:perturbationtheory}}
\end{table}

In the $m_{DM}\ll 1$ MeV regime, it is instructive to further expand $S(q,\omega)$ in the low $q$ limit, as this allows for a qualitative comparison between different channels and materials. The resulting scaling is represented schematically in Table~\ref{tab:perturbationtheory}. For a single acoustic mode in the final state, $S(q,\omega)$ scales linearly with $q$ and is by far the most favorable in terms of rate, but requires a very low threshold. For the single optical mode\footnote{The single optical mode scales as $q^4$ for dark matter that couples proportional to mass~\cite{Cox:2019cod}, which is the situation considered here; otherwise, it scales as $q^2$.} and the two-phonon processes, $S(q,\omega)$ scales as $\sim q^4$.   Quantitatively, we find that the rates of both two-phonon contributions are smaller than the rate for the single optical mode, whenever the latter is kinematically accessible.  Finally, it is interesting to compare crystals with superfluid helium, where the rate also scales as $\sim q^4$. (See Sec.~\ref{sec:helium}.) The phonon self-couplings in helium are however much stronger than in most crystals, and the reach therefore exceeds that of the cubic crystals we considered, under idealized experimental conditions. Our final, quantitative results are shown in Fig.~\ref{fig:reachplots}.
 
The paper is structured in the following way. In Section~\ref{sec:structure-factor}, we first give general expressions for the crystal structure factors that determine the rate into one and two phonons (which could be evaluated with numerical phonon eigenmodes and couplings in full generality, without the following approximations);  we further present formulae for the case of scattering in the long wavelength limit, relevant for light dark matter. In Section~\ref{sec:istropic} we introduce the isotropic approximation, and detail the elasticity theory used to determine the necessary sound speeds and anharmonic parameters for the acoustic phonons. In Section~\ref{sec:results} we present sensitivity curves for various crystals and compare rates for single and two acoustic phonon final states. We provide an estimate of rates for two-phonon final states with one or two optical phonons in Section~\ref{sec:otherchannels}, finding that they are subdominant to the single optical channel.  We also briefly comment on the qualitative differences with superfluid helium. Section~\ref{sec:discuss} concludes with a short discussion. We include three appendices which detail more lengthy aspects of the calculations, where we aim for our results to be self-contained within this paper, and reproducible.

\section{Scattering formalism}
\label{sec:structure-factor}

The scattering rate for an incident DM particle to excite phonons in a crystal is given by the dynamic structure factor, or simply structure factor.
In this section, we establish our notation and provide a derivation of the structure factors for single- and two-phonon excitations. In both cases, we obtain approximate formulae for the structure factors when the final states consist of long-wavelength acoustic phonons. In this limit, the acoustic modes have nearly linear dispersion and the structure factor can be expressed in terms of bulk properties such as  sound speeds, target density, and so on.

We begin with the most general form of the potential seen by an incident DM particle of mass $m_{DM}$:
\begin{equation} \label{eq:DM-potential}
    \mathcal{V}(\bfr) = \sum_{J=1}^{N\times \mathfrak{n}} b_J \, F({\bfr_J - \bfr}) \qquad\rightarrow\qquad  \mathcal{\tilde V}(\bfq) = \tilde F(\bfq) \sum_{J=1}^{N\times \mathfrak{n}} b_J \, e^{i \bfq \cdot \bfr_J} \,,
\end{equation}
where the index $J$ sums over all scattering centers (ions) with crystal position coordinate $\bfr_J$, $b_J$ is a factor that depends on the DM coupling with atom $J$, and the tilde indicates the Fourier-transformed function. Note that we assume from the start a crystal lattice containing $N$ primitive unit cells and $\mathfrak{n}$ ions per unit cell. Boldface symbols indicate 3-vectors in position or momentum space, while non-boldface symbols indicate scalar quantities (e.g. $q\equiv|\bfq|$). 

Two specific cases of Eq.~\eqref{eq:DM-potential} are of particular interest: a contact interaction between DM and nuclei for which $\tilde F(\bfq)=1$, and scattering via a massless mediator with $\tilde F(\bfq)\propto1/q^2$. The DM wavelength is always much larger than the radii of the nuclei, so we set the nuclear form factors to 1 everywhere.
We also assume a coupling proportional to atomic mass number, $A_J$. We then have $b_J=2\pi b_n A_J/m_{DM}$, where $b_n$ is the DM-nucleon scattering length and $\sigma_n\equiv 4\pi b_n^2$ is the DM-nucleon scattering cross section.

Pulling out the overall factor of $2\pi b_n\tilde F(\bfq) /m_{DM} $, we focus on characterizing the expectation value of the sum over scattering centers and define a dynamical structure factor given by
\begin{equation}
S(\bfq,\omega) \equiv \frac{1}{N}\sum_f\left|\sum_{J=1}^{N\times \mathfrak{n}} A_J \bra \Phi_f |e^{i\bfq \cdot \bfr_J} |0\ket\right|^2 \delta(E_f -\omega) \,,
\label{eq:Sqw_def}
\end{equation}
where $\omega$ and $\bfq$ are respectively the energy and momentum transferred from the DM to the crystal. $\bra \Phi_f|$ represents the collection of final states, indexed by $f$ and having energy $E_f$. We have assumed that the system is in its ground state $|0\ket$ before the collision; this is an excellent approximation since any dark matter experiment relying on phonons would necessarily be operating at very low temperatures, with negligibly small numbers of thermal phonons present.  Each term  then represents a scattering probability to excite a given final state. The differential cross section is moreover closely related to the structure factor; for example, taking the isotropic limit for a material,
\begin{equation}
	\frac{d^2 \sigma}{dq d\omega} = \frac{q}{2v^2 m_{DM}^2} \sigma_n  |\tilde F(\bfq)|^2 S(q, \omega) \,,
\end{equation}
where $v$ is the initial DM speed in the lab frame.

To evaluate \eqref{eq:Sqw_def} for final states with a specific number of phonons, we must expand the position vectors $\bfr_J$ in terms of equilibrium positions and displacement vectors. For a crystal with repeating primitive cells, the sum over atoms $J$ can be broken up into a sum over the lattice vectors for the primitive cells, indexed by $\bfl$, and the atoms in the primitive cell, indexed by $d$. The position operator can then be written as $\bfr_J=\bfl + \bfr^0_d +\bfu_{\bfl,d}$, where $\bfr^0_d$ is the equilibrium location of atom $d$ relative to the origin of the primitive cell, and $\bfu_{\bfl,d}$ is the displacement of that atom relative to its equilibrium position. 

We quantize the displacement vector in the harmonic approximation, following the convention in Ref.~\cite{Griffin:2018bjn}, here adapted to the Schr\"odinger picture operator:
\begin{equation}\label{eq:udefinition}
\bfu_{\bfl,d} = \sum_{\nu}^{3 \mathfrak{n}}\sum_{\bfk}  \sqrt{\frac{1}{2 N m_d \omega_{\nu,\bfk}}} \left(\mathbf{e}_{\nu,d,\bfk} \hat a_{\nu,\bfk} e^{i\bfk \cdot (\bfl+ \bfr^0_d )}+\mathbf{e}^\ast_{\nu,d,\bfk} \hat a^\dagger_{\nu,\bfk} e^{-i\bfk\cdot(\bfl+ \bfr^0_d )}  \right) \,,
\end{equation}
where there are $3\mathfrak{n}$ phonon branches, indexed by $\nu$, for a primitive cell containing $\mathfrak{n}$ atoms. Here $m_d$ is the mass of atom $d$, $\hat a^\dagger_{\nu,\bfk}$ and $\hat a_{\nu,\bfk}$ are the creation and annihilation operators for the phonons, $\omega_{\nu,\bfk}$ is the energy of phonon branch $\nu$ at momentum $\bfk$, and $\mathbf{e}_{\nu,d,\bfk}$ is the phonon eigenvector (normalized within a unit cell) for atom $d$.

Using \eqref{eq:udefinition}, the structure factor can then be expressed as
\begin{equation} \label{eq:Sqw-unitcell}
S(\bfq,\omega) = \frac{1}{N} \sum_f\left|\sum_d^\mathfrak{n} A_d e^{-W_d(0)} \mathcal{M}_{f,\bfq,d}\right|^2 \delta(E_f -\omega) \,
\end{equation}
where $W_d(0)$ is the zero-temperature Debye-Waller factor for atom $d$, and $\mathcal{M}_{f,\bfq,d}$ is the matrix element associated with final state $f$,
\begin{equation}
\label{eq:Mdefn}
\mathcal{M}_{f,\bfq,d} \equiv \sum_{\bfl} e^{i \bfq \cdot (\bfl + \bfr_d^{0})} \Big\bra \Phi_f\Big|\exp\left[ i \sum_{\bfk,\nu} \frac{\bfq\cdot \mathbf{e}^*_{\nu,d,\bfk}  }{\sqrt{2 N m_d \omega_{\nu,\bfk}}} \hat a_{\nu,\bfk}^\dagger  e^{-i\bfk\cdot(\bfl+ \bfr^0_d )} \right]\Big| 0\Big\ket \,.
\end{equation}
This expression represents the  matrix element for scattering into the crystal final state labeled by $f$, at leading order in $\widetilde{V}(q)$; however, it is not yet practical for concrete calculations. As explained in the introduction, $q/\sqrt{m_d \omega_{\nu,\bfk}}<1$ for the DM mass range of interest, which means we can consistently expand the exponential factor. 
This amounts to an expansion in the number of  phonons coupling to the DM, where the quadratic (two phonon) contribution is represented by the left-hand diagram in Fig.~\ref{fig:diagrams}. Once crystal anharmonicity is included, we also expand in the phonon self-interactions.
The leading contribution in terms of the phonon self-couplings is shown in the right-hand diagram in Fig.~\ref{fig:diagrams}. In summary, the calculation amounts to a double expansion in the momentum transfer $q$ and the phonon self-couplings.

\subsection{Single-phonon structure factor}

For the final state consisting of a single phonon with polarization $\nu$ and momentum $\bfk$, the leading result for the matrix element is 
\begin{equation}
\mathcal{M}^{(1-ph)}_{f,\bfq,d} = \sum_{\bfG} \delta_{\bfG, \bfq-\bfk} \, \frac{ i \sqrt{N} \bfq\cdot \mathbf{e}^*_{\nu,d,\bfq}}{\sqrt{2  m_d \omega_{\nu,\bfq}}}   \,   e^{i(\bfq - \bfk) \cdot \bfr^0_d} \ ,
 \label{eq:1phonon_matrixelement}
\end{equation}
where $\bfG$ are the reciprocal lattice vectors, which satisfy $\sum_{\bfl} e^{i \bfl \cdot (\bfq - \bfk)} = N \sum_{\bfG} \delta_{\bfG, \bfq - \bfk}$, with the Kronecker-$\delta$ enforcing momentum conservation in the crystal. Here we also used that phonon observables such as $\omega_{\nu, \bfq}$ are invariant under $\bfq \to \bfq + \bfG$. 
While there can be anharmonic corrections to the above matrix element, they are negligible in the low $q$ limit.

Summing over all possible single-phonon final states, this gives a structure factor identical to the result in Ref.~\cite{Griffin:2018bjn}. 
For sub-MeV DM scattering, where $q$ is typically well within the first Brillouin zone, it is a good approximation to neglect the sum over $\bfG$ as well as the Debye-Waller factors.  Then the result simplifies to 
\begin{equation}
    S^{(1-ph)}(\bfq, \omega) =  \sum_{\nu} \frac{1}{2 \omega_{\nu,\bfq}} \left| \sum_d^{\mathfrak{n}} \frac{A_d}{\sqrt{m_d}} \bfq \cdot \bfe^*_{\nu, d, \bfq} \right|^2 \delta(\omega - \omega_{\nu, \bfq}) \,.
    \label{eq:Sqw1phonon}
\end{equation}

In the long wavelength (low $q$) limit,  we can moreover approximate the acoustic modes as having real eigenvectors with magnitudes given by $|\mathbf{e}_{\nu,d,\bfq}|\approx \sqrt{m_d/(\sum^\mathfrak{n}_{d'} m_{d'})}$ and with polarization vector independent of $d$.  It is therefore convenient to introduce ``long-wavelength polarization vectors'' with unit length by defining the (real) vector
\begin{equation}\label{eq:reducedpolarization}
\bfe_{\nu,\bfq} \equiv \frac{\mathbf{e}^*_{\nu,d,\bfq}}{|\mathbf{e}_{\nu,d,\bfq}|}  \,.
\end{equation}
The difference between the two objects should be clear from the presence/absence of the index $d$ labelling atoms in the primitive unit cell. We can then simplify sums over atoms, for example by making the replacement $\sum_d A_d/\sqrt{m_d} \,  \bfe^*_{\nu,d,\bfq} \to  \bfe_{\nu, \bfq} \frac{\sum_d A_d}{\sqrt{\sum_d m_d}} \to \bfe_{\nu, \bfq} \sqrt{(\sum_d A_d)/m_p}$. In the last step, we have made the approximation that the bound atom masses are given by $m_d \approx A_d m_p$, with $m_p$ the proton mass.

As can be seen from 
\eqref{eq:1phonon_matrixelement}, transverse polarizations cannot contribute to the single phonon rate. Considering the one-phonon structure factor for longitudinal acoustic (LA) phonons,  we can take $\bfe_{\nu, \bfq} = \bfq/|\bfq|$ with the result
\begin{equation}
S^{(1-ph, LA)}(\bfq,\omega) \approx \frac{(\sum_d A_d)q^2}{2m_p \omega_{LA,\bfq}}\delta(\omega - \omega_{LA,\bfq}) \,.
\end{equation}
The LA dispersion in the long-wavelength limit is linear with slope given by the sound speed associated with the LA mode, $c_{LA}(\bfq)$, which in general can depend on the phonon propagation direction. 
Note that the factor of $\sum_d A_d$ will drop out in the expression of the rate per unit target mass, so that the rate to excite a single acoustic phonon depends only on the sound speed.\footnote{For the fiducial rate for an experiment with a real-life threshold a high sound speed  is likely more advantageous.}

\subsection{Two-phonon structure factor}

For the case with two phonons in the final state, there are two pieces which contribute to the matrix element: a contact term from expanding the exponential in \eqref{eq:Mdefn} to second order, and a piece resulting from anharmonic phonon interactions in the material.  We define $\delta H$ as the leading order anharmonic phonon interaction Hamiltonian; its precise definition we defer to Sec.~\ref{sec:anharmonic}. At leading order, the 2-phonon matrix element is then

\begin{align}\label{eq:amplitudebreakup}
\mathcal{M}^{(2\text{-}ph)}_{f,\bfq,d} &= \mathcal{M}^{(cont)}_{f,\bfq,d} +\mathcal{M}^{(anh)}_{f,\bfq,d} \,, \\
 \intertext{with}
 \mathcal{M}^{(cont)}_{f,\bfq,d} &=\sum_{\bfl} -\frac{1}{2}\Big\bra \nu_1, \bfk_1; \nu_2, \bfk_2\Big| \left[\sum_{\nu,\bfk} \frac{\bfq\cdot \mathbf{e}^*_{\nu,d,\bfk}  }{\sqrt{2 N m_d \omega_{\nu,\bfk}}} \hat a_{\nu,\bfk}^\dagger e^{-i\bfk \cdot (\bfl + \bfr_d^0)} \right]^2\Big| 0\Big\ket \,  e^{i\bfq \cdot (\bfl + \bfr_d^0)} \nonumber\\
 &= s_{1,2}\sum_{\bfG} -\frac{(\bfq\cdot \mathbf{e}^*_{\nu_1,d,\bfk_1})(\bfq\cdot \mathbf{e}^*_{\nu_2,d,\bfk_2})}{2 m_d \sqrt{\omega_{\nu_1,\bfk_1}\omega_{\nu_2,\bfk_2}}}  e^{i (\bfq - \bfk_1 - \bfk_2) \cdot \bfr_d^0} 
\, \delta_{\bfG,\bfq-\bfk_1-\bfk_2} \,,
 \label{eq:contact}\\
 \mathcal{M}^{(anh)}_{f,\bfq,d} &= i\sum_{\bfG,\bfk,\nu} \sqrt{\frac{N}{2 m_d\omega_{\nu,\bfk}}} \bigg( \frac{\bfq\cdot \mathbf{e}^*_{\nu,d,\bfk} \bra \nu_1, \bfk_1; \nu_2, \bfk_2|\delta H|\nu,\bfk\ket}{\omega_{\nu_1,\bfk_1}+\omega_{\nu_2,\bfk_2}-\omega_{\nu,\bfk} + i\Gamma_{\nu,\bfk}/2} \,  e^{i (\bfq-\bfk) \cdot \bfr_d^0}
 \, \delta_{\bfG,\bfq-\bfk} \notag \\
 &+ \frac{\bfq\cdot \mathbf{e}_{\nu,d,\bfk} \bra \nu, \bfk; \nu_1, \bfk_1; \nu_2, \bfk_2|\delta H|0\ket}{-(\omega_{\nu_1,\bfk_1}+\omega_{\nu_2,\bfk_2})-\omega_{\nu,\bfk} + i\Gamma_{\nu,\bfk}/2} \,  e^{i (\bfq+\bfk) \cdot \bfr_d^0}
 \, \delta_{\bfG,\bfq+\bfk} \bigg) \,, 
 \label{eq:anharmonicM}
\end{align}
where the factor $s_{1,2}\equiv(\delta_{\nu_1,\nu_2}\delta_{\bfk_1,\bfk_2}+1)^{-1/2}$ accounts for Bose statistics.
The contributions in \eqref{eq:contact} and \eqref{eq:anharmonicM} were shown diagrammatically in Fig.~\ref{fig:diagrams}, and  we refer to them as the \emph{contact term} and the \emph{anharmonic term}, respectively. 
Anharmonic phonon interactions also lead to a non-zero phonon width, $\Gamma_{\nu,\bfk}$. 
This has been resummed in the phonon propagator in \eqref{eq:anharmonicM} and becomes relevant when the intermediate phonon goes on-shell. 
Details regarding the derivation of the above matrix elements are given in Appendix~\ref{app:scattering-rates}.

In the long wavelength limit, we can again consider only the $\bfG=0$ contribution to the matrix elements and drop the Debye-Waller factors. 
It will then be convenient to express the three-phonon matrix element as
\begin{equation}\label{eq:expandedthreephmatrixel}
\bra \nu_1, \bfk_1; \nu_2, \bfk_2|\delta H|\nu,\bfq\ket =\frac{V}{(2 (\sum_d m_d) N)^{3/2}}\frac{\widetilde{\mathcal{M}}(\bfq,\bfk_i,\nu_i)}{\sqrt{\omega_{\nu,\bfq} \omega_{\nu_1,\bfk_1}\omega_{\nu_2,\bfk_2}}}\delta_{\bfq,\bfk_1+\bfk_2} \,,
\end{equation}
where $V$ is the volume of the crystal. As we show in Sec.~\ref{sec:anharmonic}, in the long wavelength limit  $\widetilde{\mathcal{M}}(\bfq,\bfk_i,\nu_i)$ is a function only of the momenta,  long-wavelength polarization tensors and  elastic constants of the material. 
In addition, eigenvectors are real in this limit, such that the matrix element $\widetilde{\mathcal{M}}(\bfq, \bfk_i,\nu_i)$ is real as well. The two terms in \eqref{eq:amplitudebreakup} therefore do not interfere to leading order in the small $q$ expansion, when neglecting terms higher order in $\Gamma_{\nu,\bfk}$. Using the long-wavelength polarization vectors defined in \eqref{eq:reducedpolarization}, the two-phonon structure factor can be simplified to 

\begin{align}
S(\bfq,\omega)&=S^{(cont)}(\bfq,\omega) + S^{(anh)}(\bfq,\omega) \,, \\
S^{(cont)}(\bfq,\omega)&=\frac{1}{8}\frac{\sum_d A_d}{m_p \rho}\sum_{\nu_1,\nu_2}\int\!\!\frac{d^3 \bfk_1}{(2\pi)^3}\,\frac{\left| (\bfq\cdot \mathbf{e}_{\nu_1,\bfk_1})(\bfq\cdot \mathbf{e}_{\nu_2,\bfq-\bfk_1})\right|^2}{\omega_{\nu_1,\bfk_1}\omega_{\nu_2,\bfq-\bfk_1}}\nonumber\\
&\times\delta(\omega -\omega_{\nu_1,\bfk_1}-\omega_{\nu_2,\bfq-\bfk_1})\label{eq:contactstructfun} \,, \\
S^{(anh)}(\bfq,\omega)&=\frac{1}{16}\frac{\sum_d A_d}{m_p \rho^3} \sum_{\nu_1,\nu_2}\int\!\!\frac{d^3 \bfk_1}{(2\pi)^3}\,\Bigg|
\frac{q\, \widetilde{\mathcal{M}}(\bfq,\bfk_i,\nu_i)}{\omega_{LA,\bfq}\sqrt{\omega_{\nu_1,\bfk_1}\omega_{\nu_2,\bfq-\bfk_1}}}\bigg( \frac{1}{\omega-\omega_{LA,\bfq}+i\Gamma_{LA,\bfq}/2} \nonumber \\
&+ \frac{1}{-\omega-\omega_{LA,\bfq}+i\Gamma_{LA,\bfq}/2} \bigg)\Bigg|^2 \delta(\omega -\omega_{\nu_1,\bfk_1}-\omega_{\nu_2,\bfq-\bfk_1})\label{eq:anharmstructfun} \,,
\end{align}
where we took the continuum limit by substituting $\sum_{\bfk_1}\rightarrow V \int\!\!\frac{d^3 \bfk_1}{(2\pi)^3}$. $\rho=N m_p \sum_d A_d/V$ is the mass density of the material. Similar to the single-phonon structure factor, the overall factor of $\sum_d A_d$ will drop out in the expression of the rate per unit target mass, so that the rate to excite two  phonons depends only on bulk properties such as sound speeds, density, and elastic constants.


\section{Evaluation of structure factors}
\label{sec:istropic}

In this section we provide explicit results and analytic formulae for the two contributions to the two-phonon structure factor, Eqs.~\eqref{eq:contactstructfun}-\eqref{eq:anharmstructfun}.  Even in the long-wavelength limit, the dispersions and anharmonic couplings are in general direction-dependent, substantially complicating the calculations. For cubic crystals, the isotropic limit is however known to be in excellent agreement with the general result for scattering to single phonons \cite{Griffin:2018bjn}. In this work we will therefore restrict ourselves to cubic crystals such as GaAs, Ge, Si and diamond, and approximate them as isotropic. We leave a fully general calculation of the multiphonon rate with Density Functional Theory (DFT) for future work, but we do not expect that accounting for anisotropy would qualitatively change our conclusions.

In the isotropic limit, both transverse acoustic polarizations are degenerate and the dispersion relations are simply $\omega_{LA,\bfq}=c_{LA}q$ and  $\omega_{TA,\bfq}=c_{TA}q$, with $c_{LA}$ and $c_{TA}$ the average sound speeds associated with the longitudinal acoustic (LA) and transverse acoustic (TA) modes, respectively. The structure factor for scattering to a single acoustic phonon then simplifies to
\begin{equation}
S^{(1-ph, LA)}(\bfq,\omega) \approx \frac{(\sum_d A_d)q}{2m_p c_{LA}}\delta(\omega - c_{LA} q) \,.
\label{eq:Sqw_1phonon_isotropic}
\end{equation}
For the multiphonon contribution, a description of the phonon self-interactions is needed, and this is where the isotropic approximation is most advantageous: as we will see in Sec.~\ref{sec:anharmonic}, the effective Hamiltonian is relatively simple in the isotropic and long-wavelength limit, containing 5 independent operators (this number grows to 9 if instead cubic symmetry is assumed). The coefficients of these operators can moreover be extracted from the elastic properties of the material. Each coefficient maps directly to a linear combination of the second order elastic constants (related to the bulk modulus and Young's modulus) and third order elastic constants; these quantities can either be measured or computed with \emph{ab initio} methods.

\subsection{Anharmonic term\label{sec:anharmonic}}

To compute the anharmonic contribution, we use a low-momentum effective description of the phonon self-interactions. As for any effective theory, we first constrain the form of the Hamiltonian using the symmetries of the theory and subsequently fix the Wilson coefficients from measured observables, or by matching on to the full UV theory. It is hereby convenient to introduce a ``long-wavelength displacement operator'', in analogy to the long wavelength polarization tensors defined in \eqref{eq:reducedpolarization}. Replacing the polarization tensors with their long-wavelength versions and averaging over the atoms in a unit cell, we can define
\begin{equation}\label{eq:longwavelengthu}
\bfu(\bfr) \equiv  \sum_{\nu}^{3}\sum_{\bfk}  \sqrt{\frac{1}{2 N (\sum_d m_d) \omega_{\nu,\bfk}}} \left(\mathbf{e}_{\nu,\bfk} \hat a_{\nu,\bfk}e^{i\bfk\cdot\bfr}+\mathbf{e}^\ast_{\nu,\bfk} \hat a^\dagger_{\nu,\bfk} e^{-i\bfk\cdot\bfr} \right) \,,
\end{equation}
where now we only sum over acoustic polarizations $\nu$ and we have replaced the individual atomic position vectors $\bfl + \bfr^0_{d}$ with the continuous position vector $\bfr$.
Once again, the long-wavelength displacement operators $\bfu$ can be distinguished from their more general counterparts $\bfu_{\bfl,d}$ by the index labels. 

Assuming isotropy, there are only 5 independent operators to third order in the effective Hamiltonian \cite{Tamura1984,Tamura1985}: 
\begin{equation}\label{eq:tamuraham}
\delta H=\int \!\!d^3\bfr\; \frac{1}{2}(\beta+\lambda) u_{ii}u_{jk}u_{jk}+(\gamma+\mu)u_{ij}u_{ki}u_{kj}+\frac{\alpha}{3!} u_{ii} u_{jj} u_{kk}+\frac{\beta}{2} u_{ii}u_{jk}u_{kj} +\frac{\gamma}{3}u_{ij}u_{jk}u_{ki} \,,
\end{equation}
with $u_{ij}\equiv \partial_i u_j$ and the $i,j$ running over the three spatial coordinates. Repeated indices are summed over. 
The coefficients $\alpha,\beta,\gamma,\lambda$ and $\mu$ can be determined from the measured or calculated elastic constants of the crystal.
In particular, the parameters $\mu$ and $\lambda$ are the Lam\'e parameters of the crystal and related to the bulk and Young's moduli. The parameters $\alpha$, $\beta$ and $\gamma$ can be calculated from the third order elastic constants, as described in Appendix~\ref{app:elasticity}. All five parameters have units of pressure and are reported in units of Giga-Pascal (GPa) in Tab.~\ref{tab:elastparam} for the crystals we consider.

\begin{table}[b]
\begin{ruledtabular}
\begin{tabular}{ccccccccc}
&$\mu\;\left(\text{GPa}\right)$
&$\lambda\;\left(\text{GPa}\right)$
&$\alpha\;\left(\text{GPa}\right)$
&$\beta\;\left(\text{GPa}\right)$
&$\gamma\;\left(\text{GPa}\right)$
&$c_{LA}\;\left(\text{km/s}\right)$
&$c_{TA}\;\left(\text{km/s}\right)$
&$\rho\;(\text{g/cm}^3)$\\
\hline
Si&	61&	53&	-306&	-10&	-86&	8.7 &	5.1&	2.33\\ 
GaAs&	51&	45&	-190&	-47&	-80&	5.2&	3.1&	5.32\\ 
Ge&	56&	38&	-124&	-64&	-72&	5.3&	3.2&	5.32\\ 
Diamond&	521&	86 &	-178&	-365&	-1006&	18.&	12.2&	3.51\\
\end{tabular}
\end{ruledtabular}
\caption{For a number of cubic crystals, we give the calculated elasticity parameters in the isotropic approximation, the average sound speed for the LA and TA modes, and mass density. (See Appendix~\ref{app:elasticity} for details.)  \label{tab:elastparam}}
\end{table}

Using \eqref{eq:longwavelengthu}-\eqref{eq:tamuraham}, the anharmonic three-phonon matrix element can be written in the form of \eqref{eq:expandedthreephmatrixel}, where the function $\widetilde{\mathcal{M}}$ is given by: 
\begin{align}\label{eq:tamuramatrixelem}
\widetilde{\mathcal{M}}&=
(\beta+\lambda)\Big[
(\bfq\cdot\bfe)(\bfk_1\cdot\bfk_2)(\bfe_1\cdot\bfe_2)
+(\bfk_1\cdot\bfe_1)(\bfq\cdot\bfk_2)(\bfe\cdot\bfe_2)
+(\bfk_2\cdot\bfe_2)(\bfk_1\cdot\bfq)(\bfe_1\cdot\bfe) \Big]\nonumber\\
&+(\gamma+\mu)\Big[(\bfq\cdot \bfk_1)\big[(\bfk_2\cdot \bfe_1)(\bfe_2\cdot\bfe)+(\bfk_2\cdot \bfe)(\bfe_2\cdot\bfe_1)\big]\nonumber\\
&\phantom{(\gamma+\mu)\Big[}+(\bfk_2\cdot \bfk_1)\big[(\bfq\cdot \bfe_1)(\bfe_2\cdot\bfe)+(\bfq\cdot \bfe_2)(\bfe\cdot\bfe_1)\big]\nonumber\\
&\phantom{(\gamma+\mu)\Big[}+(\bfq\cdot \bfk_2)\big[(\bfk_1\cdot \bfe_2)(\bfe_1\cdot\bfe)+(\bfk_1\cdot \bfe)(\bfe_1\cdot\bfe_2)\big]\Big]\nonumber\\
&+\alpha (\bfq\cdot \bfe) (\bfk_1\cdot \bfe_1) (\bfk_2\cdot \bfe_2)\nonumber\\
&+\beta\Big[(\bfk_1\cdot \bfe_1)(\bfq\cdot \bfe_2)(\bfk_2\cdot \bfe)
+(\bfq\cdot \bfe)(\bfk_1\cdot \bfe_2)(\bfk_2\cdot \bfe_1)
+(\bfk_2\cdot \bfe_2)(\bfq\cdot \bfe_1)(\bfk_1\cdot \bfe)\Big]\nonumber\\
&+\gamma\Big[(\bfq\cdot\bfe_1)(\bfk_1\cdot\bfe_2)(\bfk_2\cdot\bfe)+(\bfq\cdot\bfe_2)(\bfk_1\cdot\bfe)(\bfk_2\cdot\bfe_1)\Big] \,,
\end{align}
and we introduced the shorthand notation $\bfe=\bfe_{\nu,\bfq}$, $\bfe_1=\bfe_{\nu_1,\bfk_1}$ etc. From \eqref{eq:anharmonicM} it follows  that only the longitudinal polarization of the off-shell, intermediate phonon contributes. 
Depending on the polarizations of the outgoing phonons, different terms in \eqref{eq:tamuramatrixelem} contribute. Concretely, there are four distinct combinations for which the matrix element is non-zero:
\begin{itemize}
\item LA-LA
\item TA-TA with both phonons polarized in the plane spanned by the momenta
\item TA-TA with both phonons polarized orthogonal to the plane spanned  by the momenta
\item LA-TA with the TA phonon polarized in the plane spanned by the momenta.
\end{itemize}

In the isotropic limit, the structure factor in \eqref{eq:anharmstructfun} reduces to

\begin{align}
S^{(anh)}(q,\omega)&=\frac{1}{4}\frac{\sum_{d}A_d\, q^2}{\rho^3 m_p[(\omega^2-(c_{LA}q)^2)^2+(c_{LA}q)^2\Gamma_{LA,q}^2]}\nonumber\\
&\times\sum_{\nu_1,\nu_2} \int\!\!\frac{d^3 \bfk_1}{(2\pi)^3}\,\frac{|\widetilde{\mathcal{M}}|^2}{c_{\nu_1}c_{\nu_2}k_1 |\bfq-\bfk_1| }
\delta(\omega -k_1 c_{\nu_1}-|\bfq-\bfk_1| c_{\nu_2}) \,.
\label{eq:anharmstructfun2}
\end{align}
The anharmonic matrix element given in \eqref{eq:tamuramatrixelem} can also be used to compute $\Gamma_{LA,q}$, which we provide explicitly in Appendix~\ref{app:anharmform}.

The phase space integrals above can be evaluated analytically. 
Given that the different polarizations in the final states do not interfere, we can separately evaluate all four channels:

\begin{align}
S^{(anh)}_{LALA}(q,\omega)&=\frac{\sum_{d}A_d\, q^4\, \omega ^4}{16 \pi ^2 c^7_{LA} m_p \rho ^3 [(\omega^2-(c_{LA}q)^2)^2+(c_{LA}q)^2\Gamma_{LA,q}^2]} g_{LALA}^{(anh)}\left(\frac{q c_{LA}}{\omega}\right)\;\theta (\omega -c_{LA}q)\label{eq:skoAnhLALA} \,, \\
S^{(anh)}_{TATAout}(q,\omega)&=\frac{\sum_{d}A_d\, q^4\, \omega ^4}{16 \pi ^2 c^7_{TA} m_p \rho ^3 [(\omega^2-(c_{LA}q)^2)^2+(c_{LA}q)^2\Gamma_{LA,q}^2]}  g_{TATAout}^{(anh)}\left(\frac{q c_{TA}}{\omega}\right)\;\theta (\omega -c_{TA}q)\label{eq:skoAnhTATAout} \,, \\
S^{(anh)}_{TATAin}(q,\omega)&=\frac{\sum_{d}A_d\, q^4\, \omega ^4}{16 \pi ^2 c^7_{TA} m_p \rho ^3 [(\omega^2-(c_{LA}q)^2)^2+(c_{LA}q)^2\Gamma_{LA,q}^2]}  g_{TATAin}^{(anh)}\left(\frac{q c_{TA}}{\omega}\right)\;\theta (\omega -c_{TA}q)\label{eq:skoAnhTATAin} \,, \\
S^{(anh)}_{LATA}(q,\omega)&=\frac{\sum_{d}A_d\, q^4\, \omega ^4}{16 \pi ^2 c^7_{TA} m_p \rho ^3 [(\omega^2-(c_{LA}q)^2)^2+(c_{LA}q)^2\Gamma_{LA,q}^2]}  g_{LATA}^{(anh)}\left(\frac{q c_{TA}}{\omega}\right)\;\theta (\omega -c_{TA}q)\label{eq:skoAnhLATA} \,,
\end{align}
where ``TATAin'' and ``TATAout'' subscripts refer to TA-TA channels with polarizations in and orthogonal to the plane spanned by the phonon momenta. $\theta(x)$ is the Heaviside function. The $g^{(anh)}(x)$ functions all approach a constant in the $x\to 0$ limit, specifically
\begin{align}
g_{LALA}^{(anh)}\left(x\right)\approx& \  \frac{1}{240} \big(15 \alpha ^2+10 \alpha  (10 \beta +8 \gamma +5 \lambda +6 \mu )+188 \beta ^2+4 \beta  (88 \gamma +47 \lambda +66 \mu )\nonumber  \\
&+192 \gamma ^2+176 \gamma  \lambda +288 \gamma  \mu +47 \lambda ^2+132 \lambda  \mu +108 \mu ^2\big)+\mathcal{O}\left(x^2\right) \,, \\
g_{TATAout}^{(anh)}\left(x\right)\approx& \ \frac{1}{240 }\big(15 \beta ^2+10 \beta  (2 \gamma +3 \lambda +2 \mu )+12 \gamma ^2+4 \gamma  (5 \lambda +6 \mu )\nonumber\\&
+15 \lambda ^2+20 \lambda  \mu +12 \mu ^2\big)+\mathcal{O}\left(x^2\right) \,, \\
g_{TATAin}^{(anh)}\left(x\right)\approx& \ \frac{1}{16 }\big(\beta +2\gamma +\lambda +2\mu \big)^2+\mathcal{O}\left(x^2\right) \,, \\
g_{LATA}^{(anh)}\left(x\right)\approx& \ \frac{8 }{15 \delta (\delta +1)^5}(2 \beta +4 \gamma +\lambda +3 \mu )^2+\mathcal{O}\left(x^2\right) \,,
\end{align}
where we defined $\delta\equiv c_{LA}/c_{TA}$. The $\mathcal{O}(q^4)$ scaling of this contribution, as advertised in the Introduction, is therefore manifest in \eqref{eq:skoAnhLALA}, \eqref{eq:skoAnhTATAout}, \eqref{eq:skoAnhTATAin} and \eqref{eq:skoAnhLATA}. For our numerical results, we use the full, unexpanded expressions, as given in Appendix~\ref{app:anharmform}.

\subsection{Contact term\label{sec:contactterm}}

With the definition of the long-wavelength polarization tensors in \eqref{eq:reducedpolarization}, the structure factor for the contact term in \eqref{eq:contactstructfun} reduces to
\begin{equation}
S^{(cont)}(q,\omega)=\frac{1}{4}\frac{\sum_d A_d}{m_p \rho}\sum_{\nu_1,\nu_2}\int\!\!\frac{d^3 \bfk_1}{(2\pi)^3}\, \frac{\left| (\bfq\cdot \mathbf{e}_{\nu_1,\bfk_1})(\bfq\cdot \mathbf{e}_{\nu_2,\bfq-\bfk_1})\right|^2}{c_{\nu_1}c_{\nu_2}k_1 |\bfq-\bfk_1| }
\delta(\omega -k_1 c_{\nu_1}-|\bfq-\bfk_1| c_{\nu_2}) \,,
\end{equation}
which can also be evaluated analytically. Concretely, there are three final-state polarization configurations (LA-LA, TA-TA and LA-TA) which can contribute, where the TA modes must be polarized in the plane spanned by the momenta:
\begin{align}
S^{(cont)}_{LALA}(q,\omega)&=\frac{(\sum_d A_d)}{64\pi^2 c_{LA}^3 m_p\rho}\;q^4\;g_{LALA}^{(cont)}\left(\frac{c_{LA}q}{\omega}\right)\;\theta (\omega -c_{LA}q)\label{skoContLALA} \,, \\
S^{(cont)}_{TATA}(q,\omega)&=\frac{(\sum_d A_d)}{64\pi^2 c_{TA}^3 m_p\rho}\;q^4\;g_{TATA}^{(cont)}\left(\frac{c_{TA}q}{\omega}\right)\;\theta (\omega -c_{TA}q)\label{skoContTATA} \,, \\
S^{(cont)}_{LATA}(q,\omega)&=\frac{(\sum_d A_d)}{64\pi^2 c_{LA}c_{TA}(c_{LA}+c_{TA}) m_p\rho}q^4\;g_{LATA}^{(cont)}\left(\frac{c_{TA}q}{\omega}\right)\theta (\omega -c_{TA}q) \,, \label{skoContLATA}
\end{align}
with 
\begin{align}
g_{LALA}^{(cont)}\left(x\right)&\approx \frac{2}{5}-\frac{16}{21}x^2+\frac{16}{15}x^4+\mathcal{O}(x^6) \,, \\
g_{TATA}^{(cont)}\left(x\right)&\approx \frac{16}{15}-\frac{64}{35}x^2+\frac{64}{105}x^4+\mathcal{O}(x^6) \,, \\
g_{LATA}^{(cont)}(x)&\approx\frac{16}{15}+\frac{16}{105} \left(12 \delta ^2+17 \delta +5\right) x^2-\frac{16}{105} \left(7 \delta ^3+11 \delta ^2+4 \delta \right) x^4+\mathcal{O}(x^6) \,, \label{eq:contactLATAexpansion}
\end{align}
where we again used $\delta=\frac{c_{LA}}{c_{TA}}$. Note that the expansion in \eqref{eq:contactLATAexpansion} assumes $x\ll\frac{1}{\delta}$. The exact expressions for $g_{LALA}^{(cont)}$, $g_{TATA}^{(cont)}$ and $g_{LATA}^{(cont)}$ were used for all our numerical results  (see Appendix~\ref{app:contact}).  Note that the $\mathcal{O}(q^4)$ scaling discussed in the Introduction is manifest in \eqref{skoContLALA}, \eqref{skoContTATA} and \eqref{skoContLATA}. 

\subsection{Numerical comparison\label{sec:numericcomp}}
\begin{figure}
\includegraphics[height=5.8cm,trim=0.5cm 0 0 0]{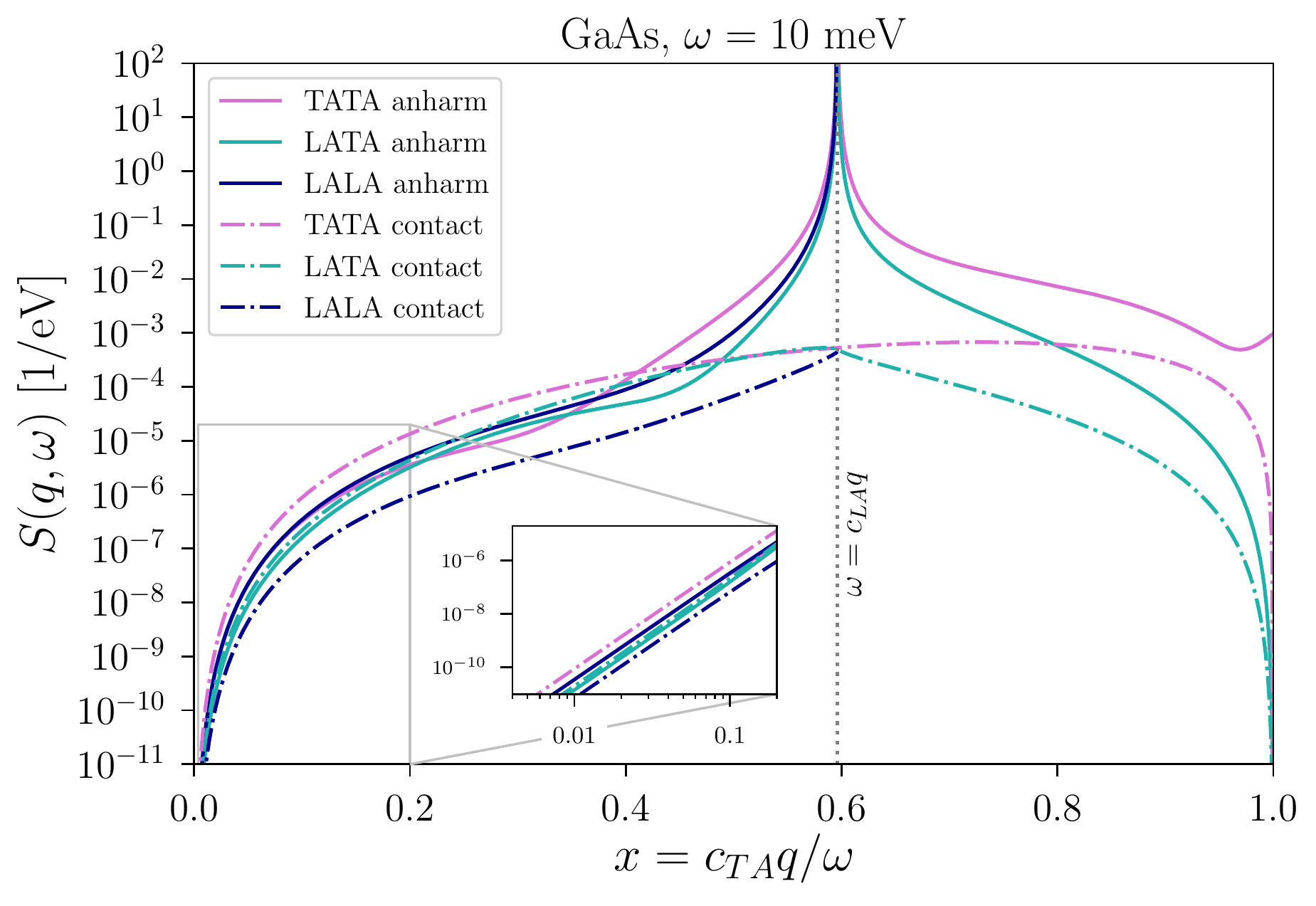}\label{fig:skoall}\hfill
\includegraphics[height=5.8cm,trim=0 -0.4cm 0 -0.1cm]{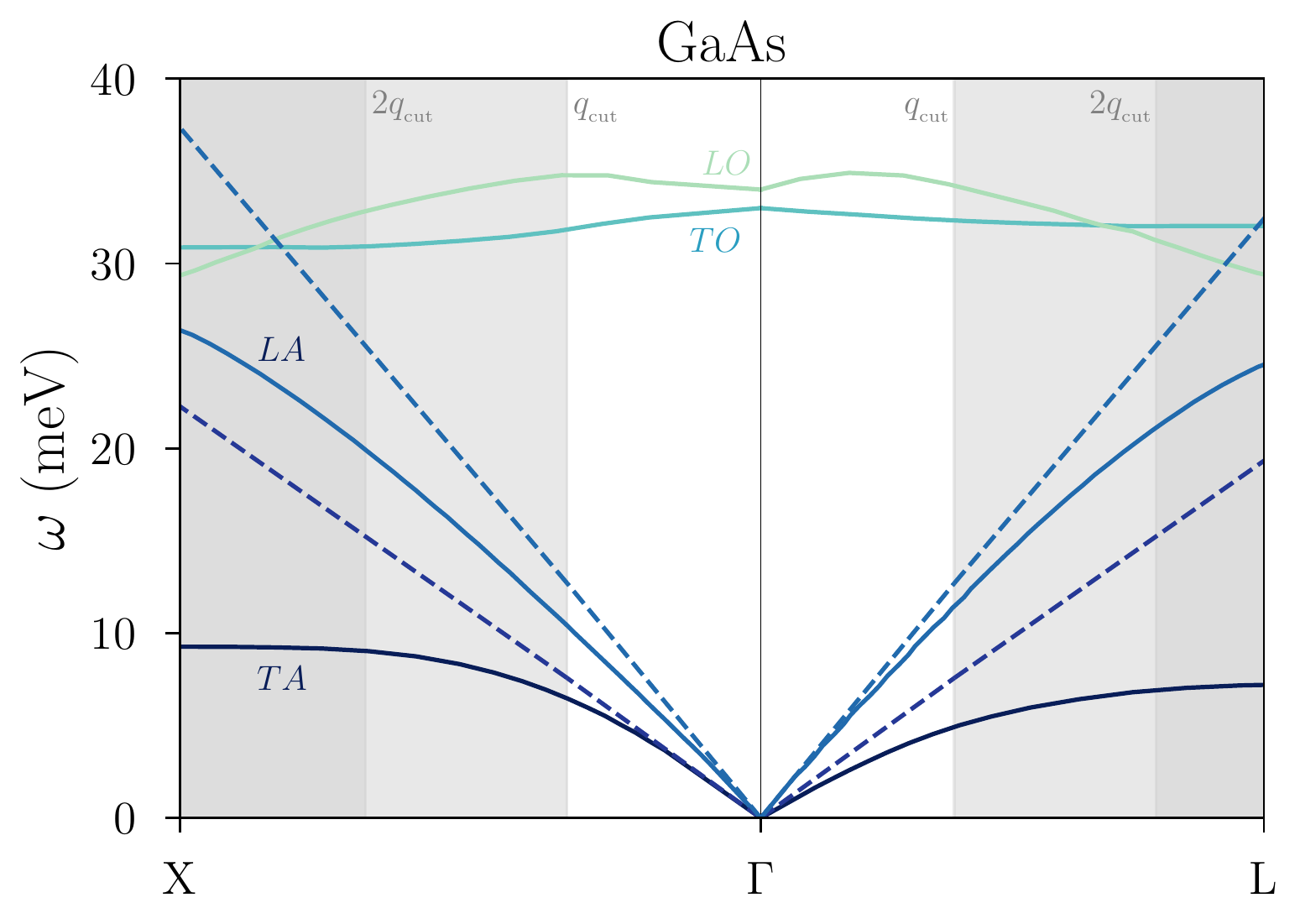}
\caption{Left: Structure factors at $\omega = 10$ meV  for each of the anharmonic and contact channels, evaluated numerically for GaAs with the parameters listed in Tab.~\ref{tab:elastparam}.   The inset shows the low-momentum regime on a log-log scale. Right: Dispersion relations for GaAs obtained with DFT methods \cite{Jain2013}, in two example directions around the origin of the Brillouin zone, indicated by ``$\Gamma$''. ($\Gamma$ in this context is not to be confused with the phonon width.) The dashed lines indicate the long wavelength, isotropic approximation, and the light and dark gray regions show $q>q_{\text{cut}}$ and $q>2q_{\text{cut}}$ respectively.
  \label{fig:skocomparison}}
\end{figure}

The left-hand panel of Fig.~\ref{fig:skocomparison} shows the different contributions to $S(q,\omega)$ for the example of GaAs, where we summed the different TATA contributions. We show the full kinematic range, where the most striking feature is the resonance at $x=1/\delta$ for the anharmonic contributions, indicating that the intermediate LA phonon goes on-shell. Whenever the resonance is kinematically accessible, it dominates the rate to the extent that the off-shell multiphonon contribution is completely negligible. The LALA channel also cuts off for $x>1/\delta$, since in this regime it is not possible to simultaneously conserve energy and momentum. Except for the region near the resonance, all contributions scale as $\omega^4$ with respect to our 10\,meV reference value. The inset zooms in on the low momentum region and shows the $\sim q^4$ scaling of both multiphonon contributions.
 
The long wavelength approximation necessarily breaks down at momenta approaching the edge of the Brillouin zone for two reasons: the dispersion relations of the acoustic phonons cease to be linear, and the description of the phonon self-couplings in terms of the elasticity parameters (Sec.~\ref{sec:anharmonic}) starts to break down. We show the dispersions in the right-hand panel of Fig.~\ref{fig:skocomparison} for the example of GaAs, where the full dispersion relations \cite{Jain2013} are compared with those in the long wavelength, isotropic approximation.  
To ensure that the calculation is not extrapolated beyond its regime of validity, we impose a maximum momentum cutoff of $q_{\text{cut}}=0.7$ keV for GaAs and Ge, $q_{\text{cut}}=0.8$ keV for Si, and $q_{\text{cut}}=1.2$ keV for diamond. This corresponds roughly to $q_{\text{cut}} \approx q_{\rm BZ}/3$, where $q_{\rm BZ} \equiv 2\pi/a$ is the approximate boundary of the first Brillouin zone and $a$ is the lattice spacing. The cut is indicated by the light gray shading in Fig.~\ref{fig:skocomparison}, and below this value the dispersions of the acoustic phonons in all four materials is close to linear.
We also enforce this momentum cut on the final state phonons by imposing an upper bound on the total deposited energy of $\omega_{\text{cut}}=(c_1+c_2) q_{\text{cut}}$ where $c_{1,2}$ stand for the sound speeds of the final state phonons under consideration, e.g.~for the LATA channel $c_1=c_{LA}$ and $c_2=c_{TA}$ etc. The resulting values are summarized in Tab.~\ref{tab:cutoffs}.

\begin{table}[t]
\begin{ruledtabular}
\begin{tabular}{cccccc}
&$a\;\left(\text{Å}\right)$
&$q_\text{cut}\;\left(\text{keV}\right)$
&$\omega_{\text{cut, TATA}}\;\left(\text{meV}\right)$
&$\omega_{\text{cut, LATA}}\;\left(\text{meV}\right)$
&$\omega_{\text{cut, LALA}}\;\left(\text{meV}\right)$\\
\hline
Si&	5.47&	0.8&	26&	35&	44\\ 
GaAs&	5.65&	0.7&	15&	20&	25\\ 
Ge&	5.66&	0.7&	16&	21&	26\\ 
Diamond&	3.57&	1.2&	94&	117&	139\\
\end{tabular}
\end{ruledtabular}
\caption{Upper bounds on $q$ and $\omega$ used in the calculations, to ensure the validity of the long wavelength approximation. $q_{\rm cut}$ is roughly $2\pi/3a$ with $a$ the lattice spacing, and the energy cuts are calculated by imposing the momentum cut on the final state phonons. We also consider cuts that are twice the values shown here. 
\label{tab:cutoffs}}
\end{table}

Due to the relatively low sound speeds in GaAs and Ge, the phase space is substantially restricted by these consistency conditions. In this sense, our calculations should be viewed as a conservative estimate. The choice of $q_{\text{cut}}$ is to some degree arbitrary, and therefore we also compute all rates with a $q_{\text{cut}}$ that is twice the values reported in Tab.~\ref{tab:cutoffs}. This provides a measure of the sensitivity of our results to $q_{\text{cut}}$. We expect that the long wavelength formulas overestimate the structure factor when extrapolated beyond their regime of validity because of the strong growth in $q$ and $\omega$, and because the isotropic linear dispersions overestimate the mode energies at large momenta. In this sense we anticipate that the true answer is bracketed by the two cutoff choices. 
Numerically, we find that integrating fully out to the edge of the Brillouin zone does not change the rates appreciably in comparison to our upper choice of $2q_{\rm cut}\sim 2 q_\text{BZ}/3$.

\section{Results} 
\label{sec:results}
Folding in the DM velocity distribution, the total rate per unit exposure is given by
\begin{align}
	R = \frac{\sigma_n}{\sum_d A_d m_p} \frac{\rho_\chi}{m_{DM}} \int d^3 \bfv_i\, f(\bfv_i) \int^{\omega_+}_{\omega_-} \!\!\! d\omega\int_{q_-}^{q_+} \!\!\!  dq\, \frac{q}{2p_im_{DM}} |\tilde F(q)|^2 S(q, \omega) \,,
\end{align}
where $\tilde F(q)$ indicates a form factor whose functional form is determined by the properties of the particle mediating the DM-nucleon scattering process. The most common, limiting cases are $\tilde F(q)=1$ if the mediator is heavier than the DM, and $\tilde F(q)={v_0^2 m_{DM}^2}/{q^2}$ for a mediator which can be treated as massless in the scattering process. In addition, $\omega_-$ is the energy threshold of the experiment, and
\begin{equation}
q_-\equiv |p_i-p_f|,\quad q_+\equiv \text{Min}\left[p_i+p_f,q_{\text{cut}}\right], \text{ and } \omega_+\equiv\text{Min}\left[\frac{v_i^2 m_{DM}}{2}, \omega_{\text{cut}}\right] \,,
\end{equation}
where $p_i\equiv m_{DM} v_i$ and $p_f \equiv m_{DM} \sqrt{v_i^2 - 2 \omega/m_{DM}}$ are the magnitudes of the initial and final DM momenta respectively. The cuts involving $q_{\text{cut}}$ and $\omega_{\text{cut}}$ ensure that the integral is not evaluated in a regime where the long wavelength approximation is invalid, as discussed in Sec.~\ref{sec:numericcomp}. For the DM velocity distribution $f(\bfv_i)$ we use the standard truncated Maxwellian distribution in the Earth's frame:
\begin{align}
f\left(\bf{v}\right) &= \frac{1}{N_0} \exp \left[-\frac{\left(\mathbf{v}+\mathbf{v}_{e}\right)^{2}}{v_{0}^{2}}\right] \Theta\left(v_{e s c}-\left|\mathbf{v}+\mathbf{v}_{e}\right|\right) \,, \\
N_0 &= \pi^{3 / 2} v_{0}^{3}\left[\operatorname{erf}\left(\frac{v_{e s c}}{v_{0}}\right)-2 \frac{v_{e s c}}{v_{0}} \exp \left(-\frac{v^{2}_{e s c}}{v_{0}^{2}}\right)\right] \,,
\end{align}
and we take $v_0=220$ km/s, $v_\text{esc}=500$ km/s, and the Earth's average velocity to be $v_e = 240$ km/s.

\begin{figure}[t]\centering
 \includegraphics[width=0.49\textwidth]{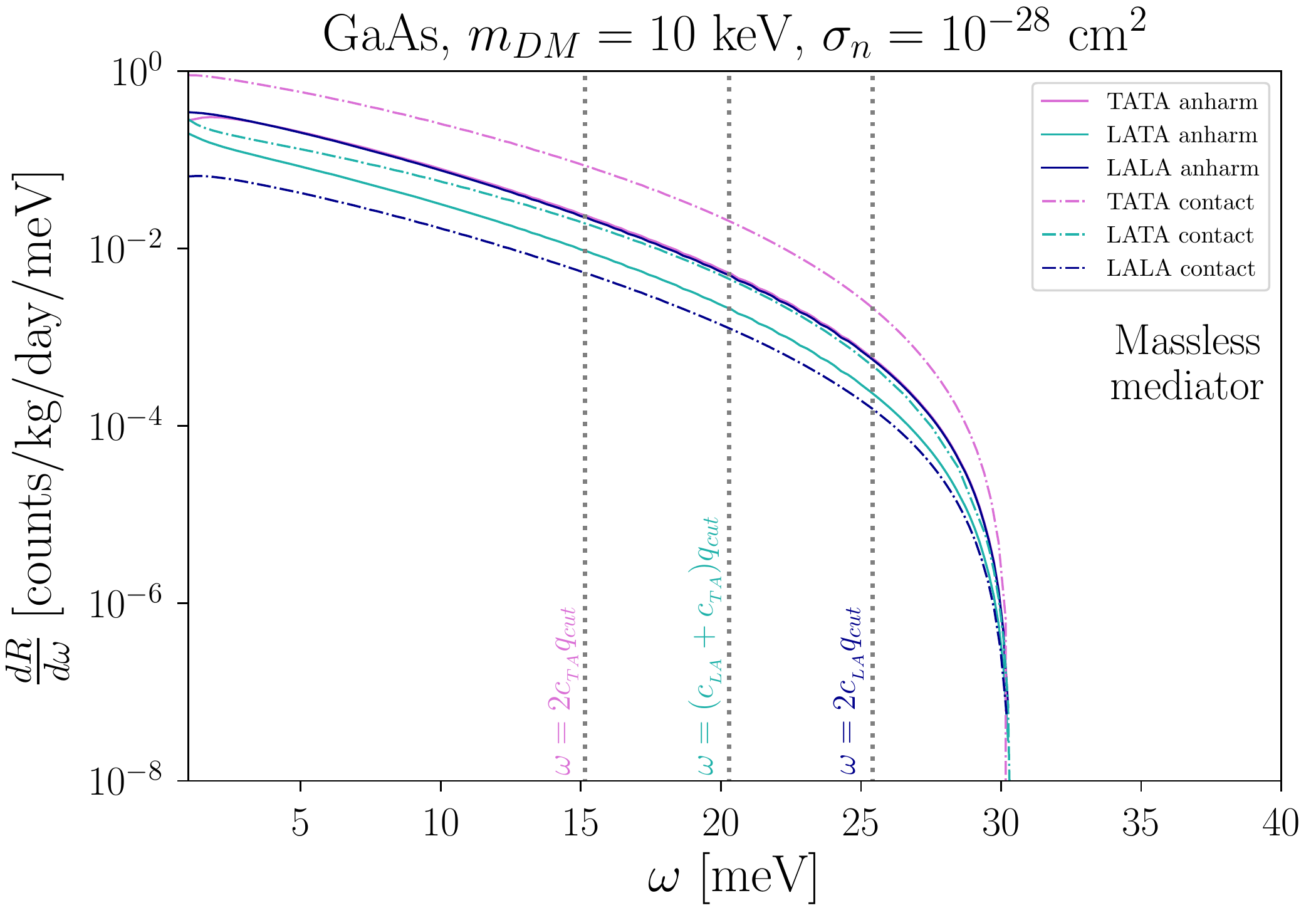}\hfill
 \includegraphics[width=0.49\textwidth]{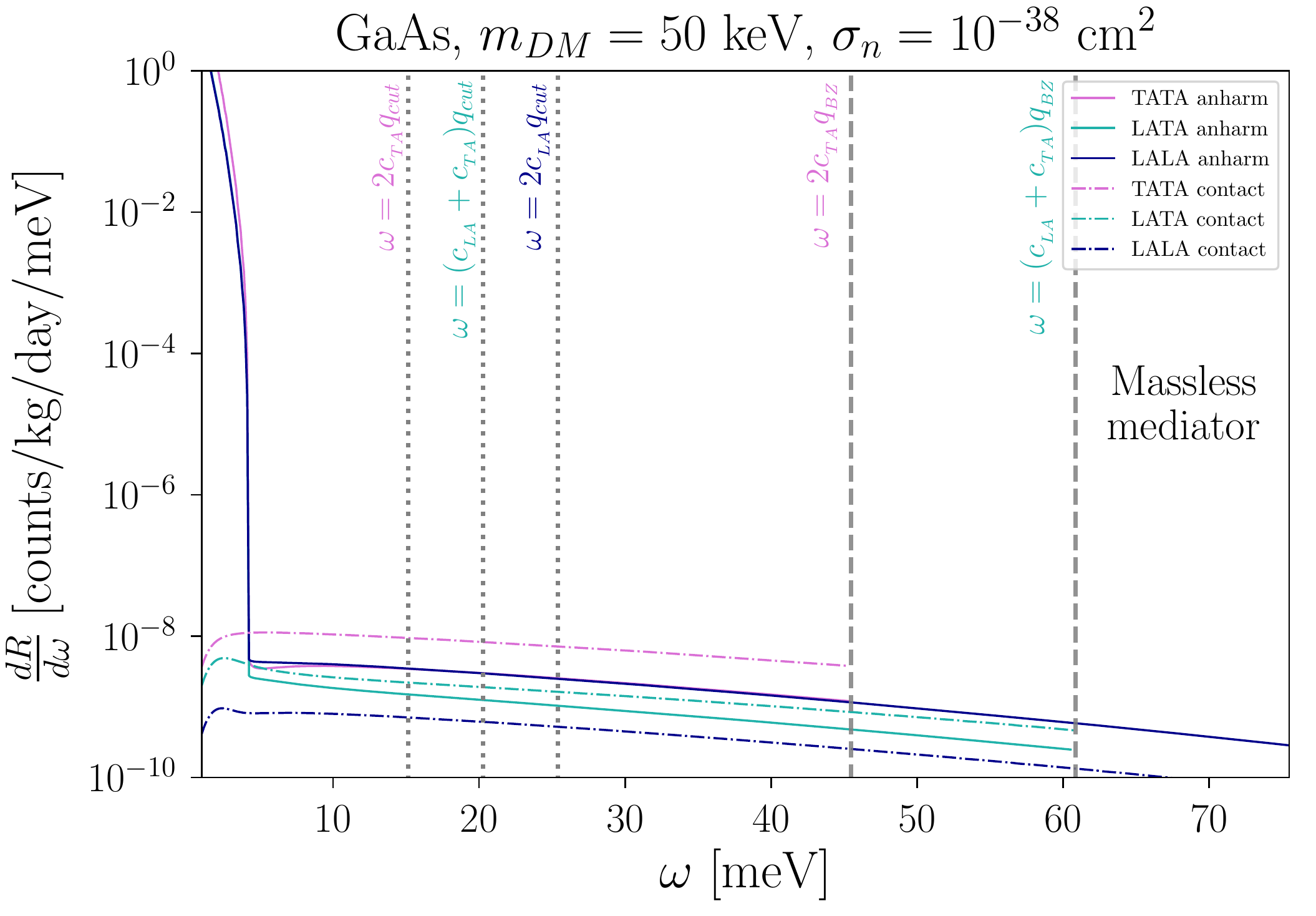}
\caption{The differential rate for the different channels in GaAs. The dotted lines indicate the $\omega$ cuts for each respective channel from  Tab.~\ref{tab:cutoffs}; the dashed lines show the cuts if we extrapolate the long wavelength approximation all the way to the edge of the Brillouin zone, and the spectra in this case should be understood as upper bounds on the true rate. The right-hand panel demonstrates the single-phonon resonance at small values of $\omega$.
\label{fig:diffrate}}
\end{figure}

Fig.~\ref{fig:diffrate} shows the differential scattering rate as a function of the deposited energy, assuming a massless mediator. 
All curves are cut off when the momenta of the final state phonons are outside the first Brillouin zone. 
The dotted vertical lines indicate values of $\omega_{\rm cut}$, above which we expect that the long wavelength approximation starts to break down. Integrating the rate beyond $\omega_{\text{cut}}$ to the edge of the Brillouin zone is likely to overshoot the true answer. 
For the $m_{DM}=10$\,keV benchmark (left-hand panel), the single phonon resonance occurs for $\omega <1$ meV, while its enormous contribution to the scattering rate is visible for $\omega <5$ meV for the 50 keV benchmark (right-hand panel). 

\begin{figure}[t]\centering
  \includegraphics[width=0.48\textwidth]{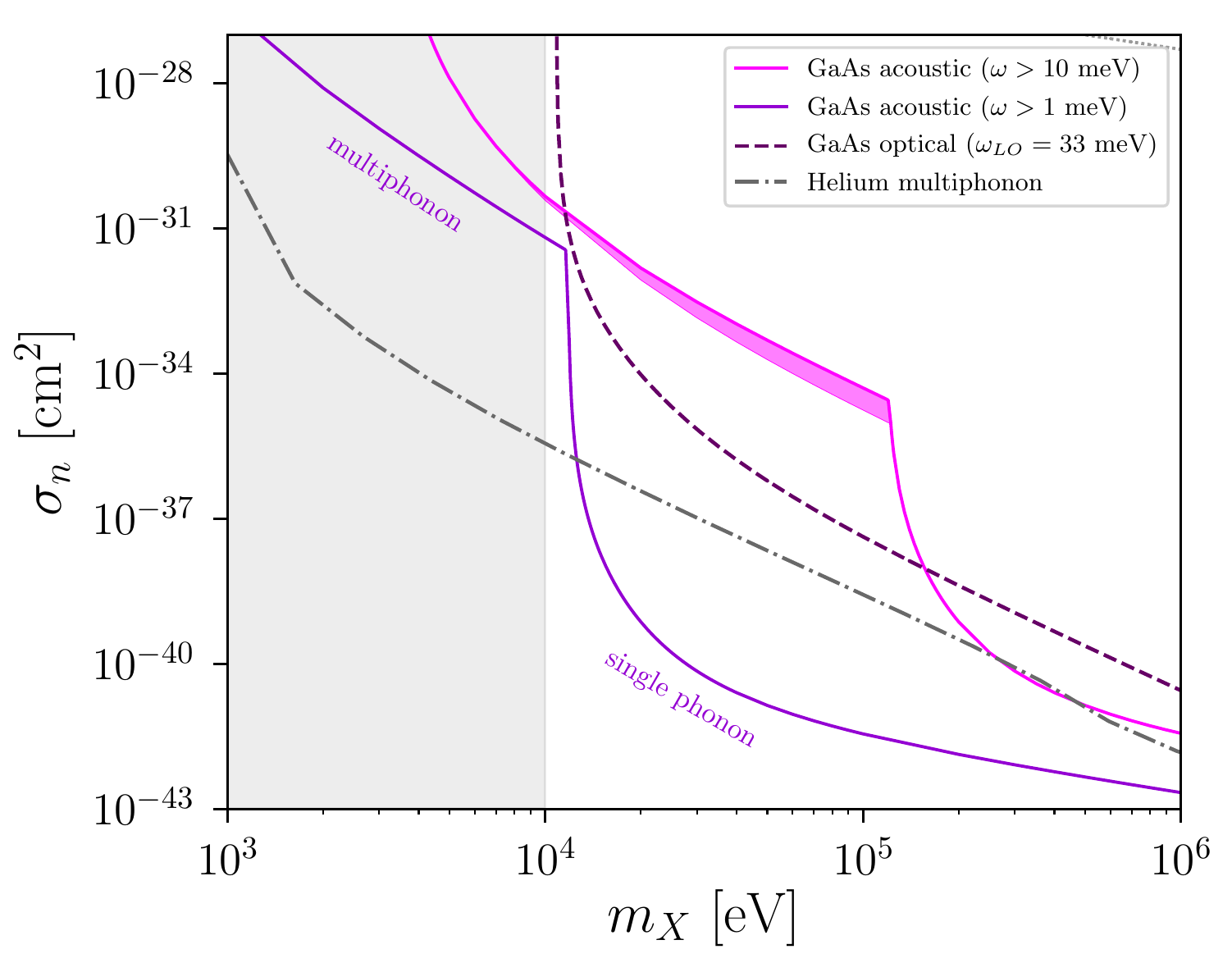}
  \includegraphics[width=0.48\textwidth]{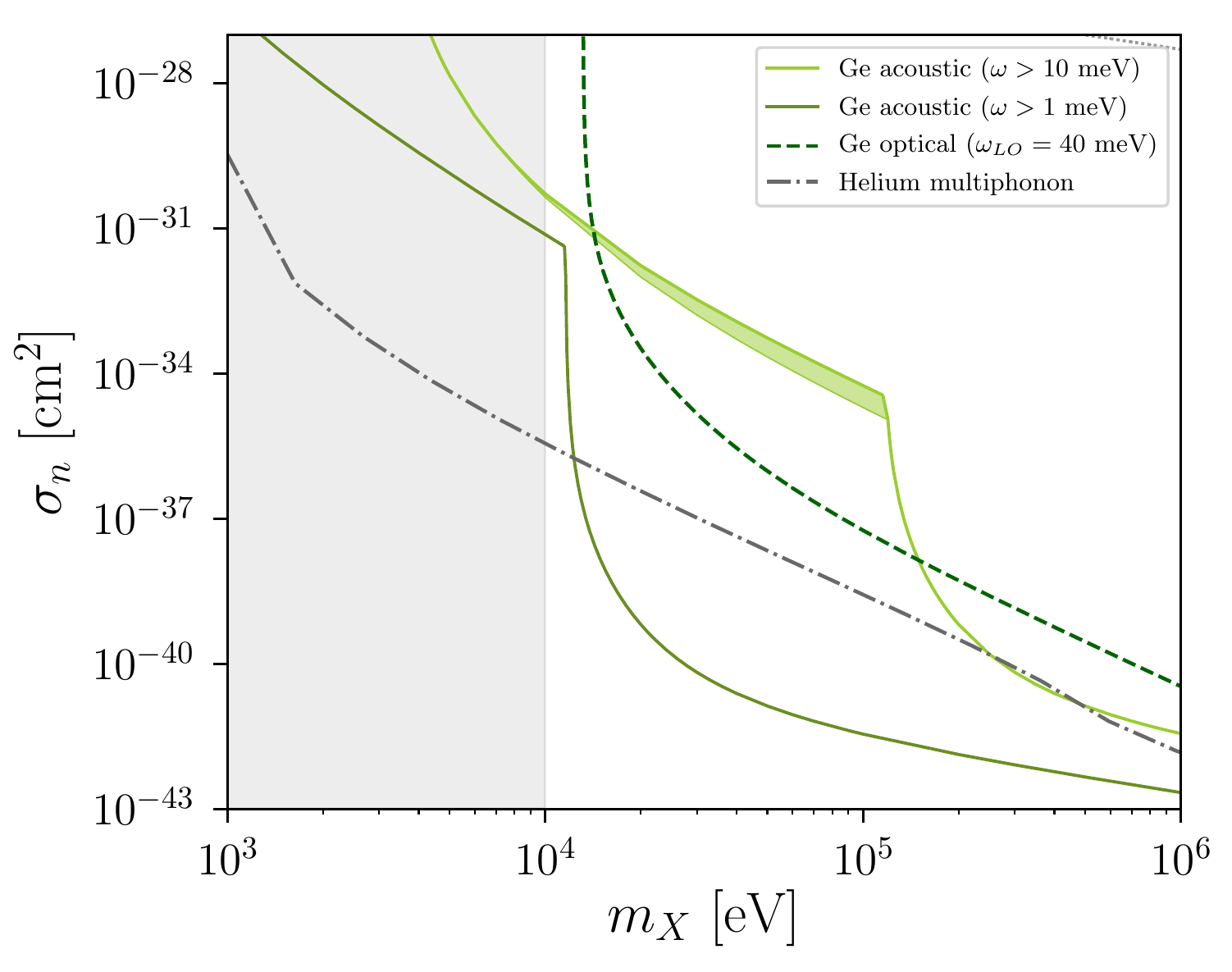}
  \includegraphics[width=0.48\textwidth]{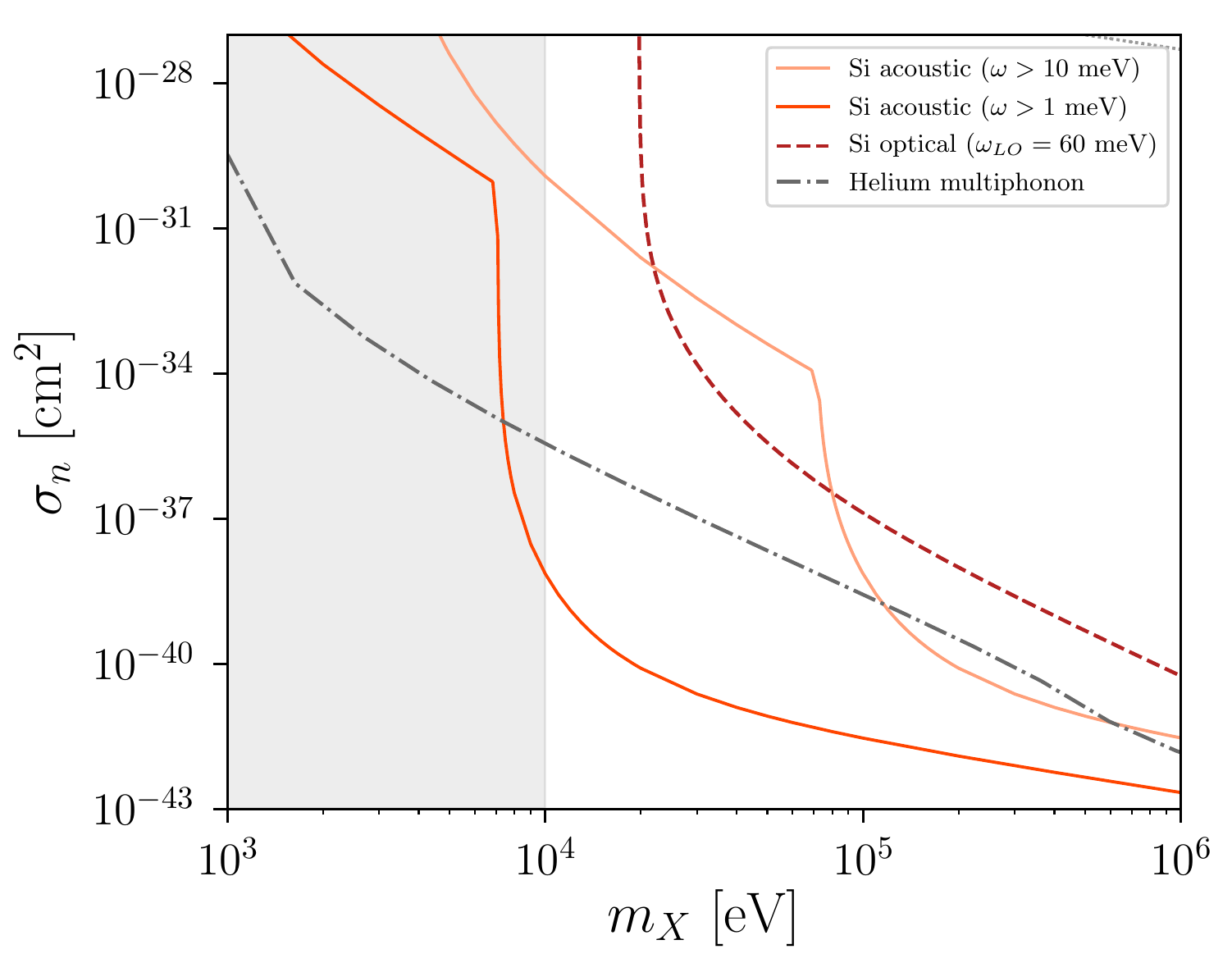}
  \includegraphics[width=0.48\textwidth]{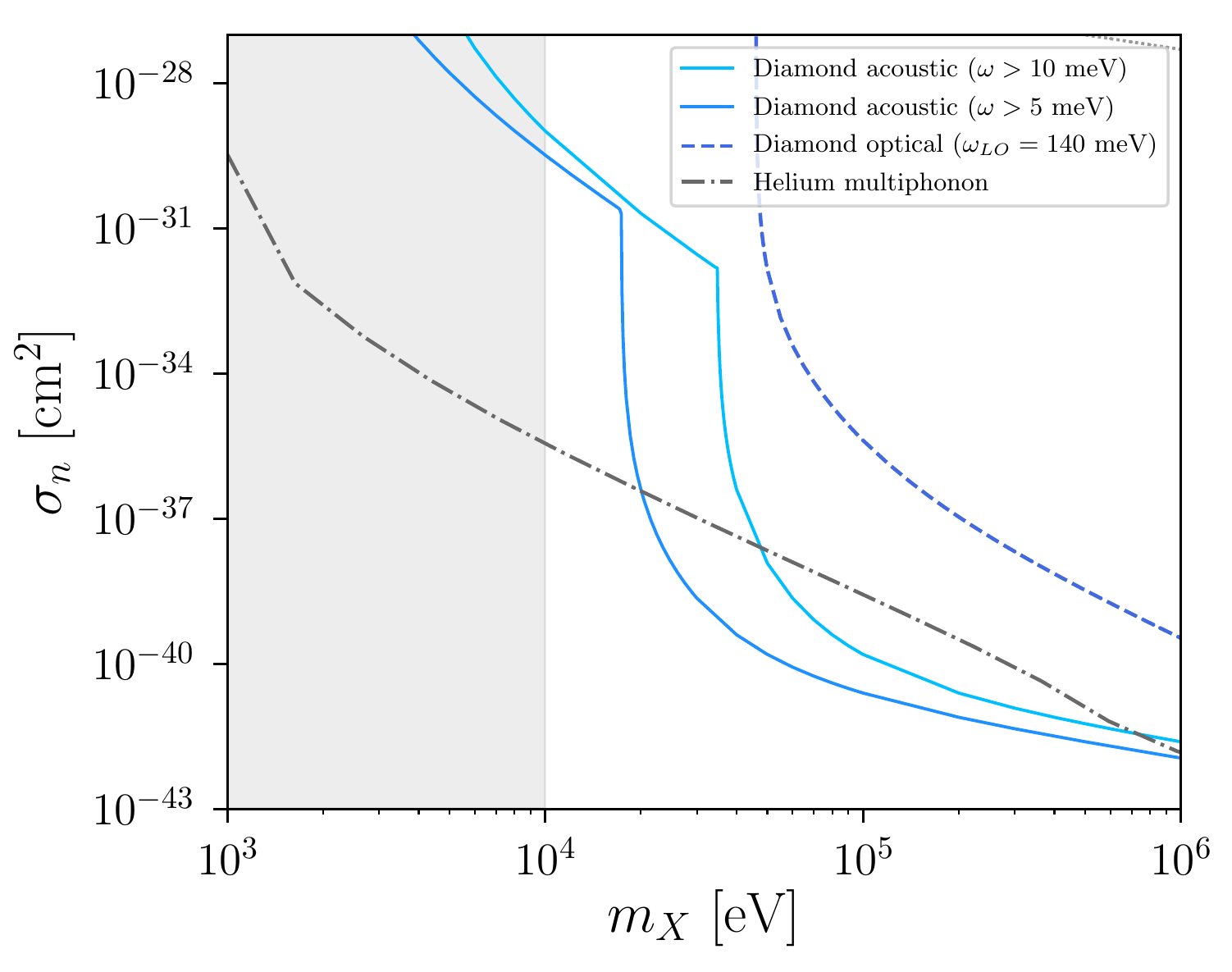}
  \caption{Minimum accessible cross sections for different crystals, channels and thresholds, assuming 3 events with a kg-year exposure.   All curves are computed in the isotropic and long-wavelength approximations. The shaded bands indicate multiphonon rates computed with the cuts in Tab.~\ref{tab:cutoffs} (upper edge) and twice those values (lower edge). The curves for the single optical channel are computed with the approximation in Sec.~\ref{sec:optical}. For comparison, we show the multiphonon reach in superfluid helium with the same exposure and a threshold of 1 meV~\cite{Knapen:2016cue}.  The dotted line in the upper right corner indicates roughly where the DM would lose a significant fraction of its initial kinetic energy within 1 km in the Earth's crust. The gray shading for $m_{DM} < 10$ keV indicates the region where stellar cooling and warm dark matter limits likely apply.
  \label{fig:reachplots}}
\end{figure}

Fig.~\ref{fig:reachplots} shows the cross sections needed to obtain 3 events with a kg-year exposure, again assuming a massless mediator.\footnote{The massive mediator scenario is disfavored by BBN bounds, while the massless mediator case is in tension with stellar cooling constraints and DM self-interactions \cite{Hochberg:2015fth,Green:2017ybv,Knapen:2017xzo}. The latter are relaxed if the particle in question is a subcomponent of the full DM abundance.} 
The most striking feature in Fig.~\ref{fig:reachplots} is the enormous enhancement of the reach once the single acoustic phonon becomes accessible.  In this regime, integrating the multiphonon structure factor matches onto the single phonon scattering rate (see Appendix~\ref{app:scattering-rates}) and we can simply use the single phonon structure factor. For a given experimental threshold $\omega_-$, the mass $m_{DM}^*$ at which the single-phonon resonance appears may be analytically derived by requiring that the maximum momentum transfer supplied by the DM suffices to create an on-shell LA phonon above the threshold, or in other words: $2 m_{DM}^*\left(  v_\text{esc} + v_e\right) \approx \omega_- / c_{LA} $ or
\begin{equation}
m_{DM}^* \approx \frac{1}{2} \frac{\omega_-}{c_{LA}} \frac{1}{\left( v_\text{esc} + v_e \right) } \,.
\end{equation}
Using the sound speed for GaAs and a $1$ meV threshold as an example, the single-phonon resonance will appear at $m_{DM}^* \approx 12$ keV, as can be seen in Fig.~\ref{fig:reachplots}.  The very high sound speed of diamond then explains why this material maintains sensitivity to the single acoustic mode for most of the mass range, even for a threshold as high as $\sim 10$ meV. (See \cite{Kurinsky:2019pgb} for a detailed study of diamond as a dark matter detector.)

No backgrounds or experimental efficiencies have been included in Fig.~\ref{fig:reachplots}, which is meant to both illustrate the most optimistic reach possible, as well as the \emph{relative} importance of the various channels, rather than provide an accurate projection of the absolute reach. The single optical phonon channel is computed using an analytic approximation given in Sec.~\ref{sec:optical}, with the (dispersionless) optical mode energy given in the figure labels. We see that the multiphonon channel is always subleading to the single optical, except at low $m_{DM}$ for Si and diamond.
The reason is that the longitudinal optical mode in both these materials is relatively high energy, 60 meV and 140 meV respectively, and is not kinematically accessible in the low $m_{DM}$ region.
For comparison, multiphonon production in superfluid helium is also shown in Fig.~\ref{fig:reachplots}; in an idealized setting where all experimental effects aside from the threshold are neglected, it always outperforms both the single optical and multiphonon modes in crystals.   

The shaded bands in Fig.~\ref{fig:reachplots} indicate the estimated uncertainty from taking the long wavelength approximation, by displaying the calculated rates using two choices for the momentum cutoff, as explained in Sec.~\ref{sec:numericcomp}. Concretely, the upper edge of the band corresponds to the values reported in Tab.~\ref{tab:cutoffs}, whereas the lower curves assume twice these values. This source of uncertainty is negligible once the single LA channel is accessible, as this contribution is peaked at low $\omega$ 
(see right-hand panel of Fig.~\ref{fig:diffrate}),
and moreover  it does not rely on the validity of Eq.~\eqref{eq:tamuraham}.
The size of the band is larger in GaAs and Ge because of the lower sound speeds and  $\omega_{\text{cut}}$ (Tab.~\ref{tab:cutoffs}). This source of uncertainty is also more severe as the experimental threshold is increased, since this reduces the available phase space in Fig.~\ref{fig:diffrate}, which leads to greater dependence on $\omega_{\text{cut}}$. For a 10\,meV threshold, 
the lower value of $\omega_{\text{cut}}$ severely restricts the phase space for the TATA channel, especially for GaAs and Ge. 
Meanwhile, for diamond $\omega_{\text{cut}}$ has no effect on the rate, since it is always larger than the initial DM kinetic energy when $m_{DM} < m_{DM}^*$. We therefore expect the long wavelength  limit  be an excellent approximation in this case.

Other sources of uncertainty are the values for the elasticity parameters, as to the best of our knowledge they have not yet all been measured at ultra low temperatures. As explained in Appendix~\ref{app:elasticity}, we instead rely on \emph{ab initio} calculations of these parameters, which in some cases carry $\mathcal{O}(1)$ uncertainties. This propagates to a small uncertainty on the overall multiphonon rate, regardless of the DM mass. In addition, we expect corrections to the isotropic approximation once the detailed crystal structure is accounted for. These uncertainties are not included in the band in Fig.~\ref{fig:reachplots}. Given the current experimental unknowns, we consider the uncertainties acceptable at this stage, especially given that multiphonon processes typically have a much lower rate than the single optical mode.

To conclude, we briefly comment on stellar cooling constraints, warm dark matter bounds and the material overburden. For millicharged particles with mass $\lesssim 10$ keV, there are strong constraints from the cooling of white dwarfs, red giants and horizontal branch stars \cite{Davidson:2000hf,Vogel:2013raa}. To our knowledge, the analogous computation has not yet been performed for light DM with a coupling to nuclei, but we expect that similar constraints should apply for $m_{DM}\lesssim 10$ keV. In this mass range, the DM is also generally considered as warm and there are constraints from structure formation, although these are alleviated if this candidate doesn't provide the entire DM abundance. The likely existence of both bounds is suggested by the gray shading in Fig.~\ref{fig:reachplots}. Finally, for sufficiently large $\sigma_n$, the DM is likely to scatter in the Earth's crust before reaching an underground detector. 
To determine roughly where this occurs, we estimated the mean free path for DM scattering off phonons in a crystalline silicon crust where the DM loses at least 1\% of its typical initial kinetic energy. (While this is an idealized model, a similar result is obtained if we model DM interactions in the crust as nuclear recoils off free silicon atoms.)  The dotted line in Fig.~\ref{fig:reachplots} indicates where the mean free path is 1 km. Numerically, we find this to be where $\sigma_n \gtrsim 5\times 10^{-28} \text{ cm}^2 \times \left(\text{MeV}/{m_{DM}} \right). $

\section{Other channels \label{sec:otherchannels}}

\subsection{Multiphonons involving optical branches\label{sec:optical}}

As discussed in the introduction, the rate for scattering that excites a single optical phonon is suppressed when the DM coupling is proportional to the mass of the atom. 
Nevertheless, as seen in the previous section, processes involving optical phonons are still important, particularly for higher experimental thresholds. 
In this section, we briefly review the single longitudinal optical (LO) phonon calculation, before discussing two-phonon processes involving optical phonons.

To obtain an estimate of the rate to excite a single LO phonon, we use an  approximation for the eigenmode in a cubic lattice with diamond or zincblende structure, valid at low $q$:
\begin{equation} \label{eq:optical-eigenvector}
    e_{LO,1,\bfq} \approx \frac{\sqrt{A_2}}{\sqrt{A_1 + A_2}} \,, \quad e_{LO,2,\bfq} \approx -\frac{\sqrt{A_1}}{\sqrt{A_1 + A_2}} e^{- i \bfq \cdot \bfr_2} \,,
\end{equation}
where $\bfr_2 = (a/4,a/4,a/4)$ is the position of the second atom in the primitive cell and $a$ is the lattice constant.
Note that without the phase factor the structure factor would be exactly zero; we have included it to account for the subleading behavior~\cite{Cox:2019cod}. Using eq.~\eqref{eq:Sqw1phonon} and averaging over angles such that $(\bfq \cdot \bfr_2)^2 \approx q^2 a^2/16$, we obtain 
\begin{equation}
    S_{LO}(q,\omega) = \frac{q^4 a^2}{32 \omega_{LO}}  \frac{A_1 A_2}{m_p (A_1 + A_2)}  \delta( \omega - \omega_{LO}) \,,
\end{equation}
where we have approximated the LO phonon's dispersion relation as flat, $\omega_{LO}(\bfq) = \omega_{LO}$. This approximation reproduces the full numerical result for the DM reach in GaAs (see Ref.~\cite{Griffin:2018bjn}) to within an O(1) factor. 

There are two kinds of two-phonon processes involving optical phonons to consider: optical-acoustic, and optical-optical.
We begin with the former, since they are the most relevant for light DM.
Optical-acoustic scattering also has both contact and anharmonic contributions. 
For all of the materials we consider, there is a suppression of the contact contribution at low $q$ when DM couples proportional to atomic mass. 
This can be seen from the expressions for the structure factor and matrix element in eqs.~\eqref{eq:Sqw-unitcell} and \eqref{eq:contact}. 
When $q=0$, momentum conservation requires $k_1=-k_2$ and the sum over the unit cell in \eqref{eq:Sqw-unitcell} vanishes due to the orthogonality of the eigenvectors. 
Using the low-$q$ approximation for the LO eigenvector \eqref{eq:optical-eigenvector}, one can explicitly see that the leading term in the small $q$ expansion of the structure factor vanishes; the contact term then scales as $q^6$ and is negligibly small. 
Note that this result does not hold for general lattices, since with more complicated unit cells there can be mixed longitudinal-transverse optical modes which may only be orthogonal to the acoustic modes after also contracting the Lorentz indices of the eigenvectors. 

The anharmonic contribution is more difficult to reliably calculate. 
It could be obtained from a first principles calculation of the anharmonic corrections to the lattice potential using Density Functional Theory, however this goes beyond the scope of the present paper. 
Here, we adopt a simpler method in order to obtain an estimate of the size of this contribution.
We follow an approach that has been used in the literature to calculate the lifetime of LO phonons and describe the anharmonic three-phonon interactions via the Hamiltonian~\cite{Srivastava:1980},
\begin{multline} \label{eq:optical-interactions}
  \delta H = \frac{1}{3!} \frac{\gamma_G}{\bar{c}} \sqrt{\frac{1}{2Nm_p(A_1+A_2)}} \sum_{\nu,\nu',\nu''} \sum_{\bfk,\bfk',\bfk''} \sqrt{\omega_\nu \omega_{\nu'} \omega_{\nu''}} \, \delta_{\bfk+\bfk'+\bfk''} \\ \times \left( a_{\nu,\bfk}^\dagger - a_{\nu,\bfk}\right) \left( a_{\nu',\bfk'}^\dagger - a_{\nu',\bfk'}\right) \left( a_{\nu'',\bfk''}^\dagger - a_{\nu'',\bfk''}\right) \,,
\end{multline}
where $\gamma_G \approx 1$ is the mode-averaged Gr\"uneisen constant and $\bar{c}$ is the average of the LA and TA  sound speeds. 
The above Hamiltonian can be obtained  starting from eqs.~\eqref{eq:longwavelengthu}-\eqref{eq:tamuramatrixelem} and then averaging over phonon modes and angles (see Ref.~\cite{Srivastava:1980}). 
Since this model treats the lattice as an isotropic continuum it does not actually contain optical modes; nevertheless,  eq.~\eqref{eq:optical-interactions} has been used in the calculation of optical phonon lifetimes (e.g.~\cite{PhysRevB.43.4939, PhysRevB.69.235208}). 

The dominant anharmonic contribution is that mediated by an off-shell LA phonon, since the LO mediated process has the same suppression as single optical scattering. 
Using the Hamiltonian \eqref{eq:optical-interactions} in eq.~\eqref{eq:anharmonicM} we obtain the structure factors,

\begin{align} 
  S_{LOLA}^{(anh)}(q,\omega) = \frac{\gamma_G^2}{2\pi^2} \frac{\omega_{LO} (A_1+A_2)}{\bar{c}^2 c_{LA} \rho m_p } \frac{q^4 (\omega - \omega_{LO})^3}{(\omega^2-(c_{LA}q)^2)^2} \theta(\omega-\omega_{LO}) \,, \label{eq:LOLA} \\
  S_{LOTA}^{(anh)}(q,\omega) = \frac{\gamma_G^2}{\pi^2} \frac{\omega_{LO} (A_1+A_2)}{\bar{c}^2 c_{TA} \rho m_p} \frac{c_{LA}^2}{c_{TA}^2} \frac{q^4 (\omega - \omega_{LO})^3}{(\omega^2-(c_{LA}q)^2)^2} \theta(\omega-\omega_{LO}) \,, \label{eq:LOTA}
\end{align}
where we have again assumed a flat dispersion relation for the optical mode. 
The expressions for the TO-LA and TO-TA processes can be obtained by the substitution $\omega_{LO}\to\omega_{TO}$ and multiplying by a factor of two.
Integrating the structure factor to obtain the total rate we find that, for all the materials we consider, 
the LO-LA scattering rate is four to five orders of magnitude smaller than the single optical rate, where we again impose the $q_{cut}$ values in Tab.~\ref{tab:cutoffs} on the acoustic phonons (relaxing this cut increases the LO-LA rate, but it always remains negligible). The LO-TA process is enhanced by the smaller TA sound speed, but is still significantly suppressed compared to the single optical. 
A similar conclusion holds for optical-acoustic scattering involving TO phonons, although these processes could be relevant in a narrow range of DM masses that are above the threshold to excite a TO phonon but below the LO threshold. 
While eqs.~\eqref{eq:LOLA} and \eqref{eq:LOTA} should only be considered as an estimate of the two phonon optical-acoustic rate, we do not expect a detailed DFT calculation to change the qualitative conclusion that it is sub-leading compared to single optical scattering.

Next, we briefly discuss scattering into two optical phonons. 
This process only becomes kinematically accessible for heavier DM masses due to the higher energy threshold to excite two optical phonons. 
Unlike optical-acoustic scattering, there is no additional suppression of the contact contribution for DM that couples proportional to atomic mass. 
The LO-LO structure factor is then proportional to $q^4/(m_p \omega_{LO})^2$.  
On the other hand, the single optical structure factor scales as $q^4a^2\mu/(m_p^2 \omega_{LO})$, where $\mu$ is the reduced mass of the primitive cell. 
The two optical phonon contact contribution is then expected to be significantly smaller than the single optical. 
The anharmonic contribution is again challenging to reliably estimate; however, based on our above estimate for optical-acoustic scattering, where it was found to be sub-leading, we do not expect it to give a significant contribution. 
In summary, two phonon scattering processes involving optical phonons are expected to give only a sub-leading contribution to the total scattering rate.

\subsection{Multiphonons in superfluid helium\label{sec:helium}}
Here we briefly compare our results with similar calculations of multiphonon production in superfluid helium. While the symmetries of the systems are different, in both cases the structure factor scales as $q^4$ in the limit $q \ll \omega$. Crystals spontaneously break both translation and rotation invariance, but since the rotation operators are linearly dependent on the translation operators, there are only 3, rather than 6, Goldstone modes \cite{Watanabe:2013iia}. These are the 1 LA and 2 TA modes we have encountered throughout our discussion. Since translations are broken spontaneously, all amplitudes must vanish in the limit where one of the external (spatial) momenta go to zero. This symmetry principle explains the form of the amplitude in \eqref{eq:tamuramatrixelem} and its scaling in the low $q$ limit. Combined with the $q$-dependent DM-phonon coupling, the resulting matrix element goes as $|\mathcal{M}|^2 \sim q^4$.


Superfluid helium on the other hand does not break translation and rotation invariance, though the Bose-Einstein condensate breaks boost invariance as well as a linear combination of the time translation and particle number operators. All four broken operators are linearly dependent, such that there only exists a single Goldstone mode \cite{Watanabe:2013iia}, which is the phonon-roton branch. 
Here the same $q^4$ dependence of the structure factor can be argued from an effective field theory treatment~\cite{Acanfora:2019con,Caputo:2019cyg}. Although translation invariance is unbroken, the Ward identity associated with the $U(1)$ particle number symmetry still enforces that the two-phonon amplitude vanishes in the $\bfq\to0$ limit \cite{esposito}.  Bose symmetry on the final state momenta then implies that in the low $q$ limit, the amplitude must be proportional to
\begin{equation}\label{eq:simplehe}
|\mathcal{M}_{\mathrm{He}}|^2 \;\sim\; \left|\bfq\cdot \bfk_1+ \bfq\cdot \bfk_2\right|^2 \sim q^4 \,,
\end{equation}
where the second $\sim$ follows from momentum conservation ($\bfq =\bfk_1+\bfk_2$). 
Despite the differences in symmetries, the scaling of the dynamic structure factor for phonons in superfluid helium is the same as for longitudinal acoustic phonons in crystals. However, the multiphonon rate  in helium exceeds that in the crystals we considered (see Fig.~\ref{fig:reachplots}), due to the stronger phonon self-couplings in helium.
\section{Conclusions and outlook}
\label{sec:discuss}
In this work, we evaluated the rate for production of two acoustic phonons in crystals from scattering of sub-MeV DM. We considered cubic crystals such as GaAs, Ge, Si and diamond and worked in the isotropic and long wavelength approximations. In addition, we focused on DM which couples proportional to atomic mass, since in this case  the rate for single optical phonon excitations is suppressed and multiphonon production is most relevant. However, for all four crystals, we found that the multiphonon rate is smaller than the single optical phonon rate  whenever the optical mode is kinematically accessible. Similarly, the rate to excite a single acoustic phonon dominates whenever that mode is kinematically accessible. In diamond and Si there is, however, a range of DM masses between 10 keV and 100 keV for which the multiphonon process could be the only detectable channel, depending on the experimental threshold. We have also estimated the multiphonon rate with optical phonons and expect it to be sub-leading. In idealized experimental conditions, the multiphonon rate in superfluid helium exceeds that in all the crystals we have considered.

For GaAs and Ge, our approach here in taking the long wavelength approximation has a limited regime of validity, leading to appreciable uncertainties in the scattering rate. A more precise evaluation with Density Functional Theory methods would be desirable for these materials. Such a DFT treatment would also allow one to study anisotropic materials such as sapphire, which are expected to exhibit a sizable daily modulation in the multiphonon signal.

\acknowledgments
We thank Christian Bauer, Angelo Esposito, Marat Freytsis, Sin\'ead Griffin, Katherine Inzani, Aneesh Manohar, Harikrishnan Ramani, Nicholas Llewellyn Rodd, Tanner Trickle, James Wells, Zhengkang Zhang and Kathryn Zurek for helpful discussions, and Angelo Esposito for comments on a draft version of the manuscript.  SK and TL thank the Munich Institute for Astro- and Particle Physics (MIAPP) and the Galileo Galilei Institute for Theoretical Physics for their hospitality and the INFN for partial support during the completion of this work. SK also thanks the Aspen Center for Physics, supported by National Science Foundation grant PHY-1607611. SK is supported by DOE grant DE-SC0009988 and the Paul Dirac fund at the Institute for Advanced Study.   TL is supported by an Alfred P. Sloan Research Fellowship and Department of Energy (DOE) grant DE-SC0019195. PC is supported by the Australian Research Council. TM is supported by JSPS KAKENHI Grant Number JP18K13533 and JSPS KAKENHI Grant Number JP19H05810. PC and TM are supported by the World Premier International Research Center Initiative (WPI), MEXT, Japan. 

\clearpage

\appendix

\section{Derivation of scattering rates} \label{app:scattering-rates}

In this appendix we show how the matrix elements in Sec.~\ref{sec:structure-factor} are derived using time-dependent perturbation theory, including a resummation of the phonon width.

We begin by rewriting the Hamiltonian as
\begin{equation}
  H = \frac{p_{DM}^2}{2m_{DM}} + \sum_{\nu,\bfk} \left( \omega_{\nu,\bfk} - \frac{i}{2} \Gamma_{\nu,\bfk} \right) a^\dagger_{\nu,\bfk} a_{\nu,\bfk} + H' \,,
\end{equation}
with
\begin{equation}
  H' = \mathcal{V}(\bfr) + \delta H + \sum_{\nu,\bfk} \frac{i}{2} \Gamma_{\nu,\bfk} a^\dagger_{\nu,\bfk} a_{\nu,\bfk} \,,
\end{equation}
where $\mathcal{V}$ and $\delta H$ are given in eqs.~\eqref{eq:DM-potential} and \eqref{eq:tamuraham} respectively, and we have introduced the phonon width, $\Gamma_{\nu,\bfk}\sim\mathcal{O}(\delta H^2)$. 
In the following, $H'$ will be treated as a perturbation. 
Introducing the phonon width in this way is purely a reorganisation of the perturbation series as the full Hamiltonian remains independent of $\Gamma_{\nu,\bfk}$. 
This approach is similar to the complex mass scheme in QFT~\cite{Denner:1999gp} and allows for a systematic inclusion of the width at higher perturbative orders, although is not strictly necessary here since we consider only the leading corrections from $\delta H$.

Using the above Hamiltonian, we calculate the dark matter scattering rate using time-dependent perturbation theory. 
We assume that the system is initially described by the $H'=0$ Hamiltonian at $t_0\to-\infty$, and adiabatically turn on the perturbation by replacing $H' \to e^{\epsilon t}\, H'$, where we eventually take the limit $\epsilon\to0$. 
Specifically, we take the initial state to be $|p_i;0\ket$, where $p_i$ is the dark matter momentum and the phonons are in the ground state.

\subsection{Single phonon}

For scattering into a single phonon, the anharmonic correction is negligible and it is sufficient to consider only the leading order contribution. 
The transition probability to scatter and be in the state $|p_f;\nu,\bfk\ket$ at some time $t$ is
\begin{equation} \label{eq:transition-prob}
  |\matrixel{p_f;\nu,\bfk}{U(t,-\infty)}{p_i;0}|^2 = \frac{\left|\matrixel{p_f;\nu,\bfk}{\mathcal{V}(\bfr)}{p_i;0}\right|^2}{(\omega_{\nu,\bfk}-\omega)^2 + (\Gamma_{\nu,\bfk}/2 + \epsilon)^2} \, e^{2\epsilon t} \,,
\end{equation}
where ${U(t,-\infty)}$ is the time evolution operator in the Schr\"odinger picture, and $|p_f;\nu,\bfk\ket$ is an eigenstate of the $H'=0$ Hamiltonian. 
For scattering into stable final states ($\Gamma_{\nu,\bfk} = 0$) the transition rate is just Fermi's Golden Rule:
\begin{equation}
  w_{i\to f} \equiv \lim_{\epsilon\to0}\frac{d}{dt} |\matrixel{p_f;\nu,\bfk}{U(t,-\infty)}{p_i;0}|^2 = 
    2\pi \delta(\omega_{\nu,\bfk}-\omega) |\matrixel{p_f;\nu,\bfk}{\mathcal{V}(\bfr)}{p_i;0}|^2 \,.
\end{equation}
Substituting in Eqs.~\eqref{eq:DM-potential} \& \eqref{eq:udefinition} this becomes
\begin{equation}
    w_{i \to f} = 2\pi \delta(\omega_{\nu,\bfk}-\omega) \left(\frac{2\pi b_n}{m_{DM}V}\right)^2 \big|\tilde{F}(q)\big|^2 \left|\sum_d^\mathfrak{n} A_d e^{-W_d(0)} \mathcal{M}^{(1-ph)}_{|\nu,\bfk\ket,\bfq,d}\right|^2 \,,
\end{equation}
where $\mathcal{M}^{(1-ph)}$ is defined in Eq.~\eqref{eq:1phonon_matrixelement}, and $V$ is a volume factor from the normalisation of the DM momentum eigenstates. The transition rate is directly related to the structure factor in Eq.~\eqref{eq:Sqw-unitcell} up to overall factors. 

Note that for unstable final states ($\Gamma_{\nu,\bfk} \neq 0$) $w_{i \to f}$ vanishes.
In this case the transition probability in Eq.~\eqref{eq:transition-prob} does not grow with time (it is constant when $\epsilon\to0$), since due to the exponential decay of the state only the last $\Delta t \sim\Gamma_{\nu,\bfk}^{-1}$ contributes significantly.

\subsection{Two phonon}

Next, consider scattering into the two phonon state $|p_f;\nu_1,{\bfk_1};\nu_2,{\bfk_2}\ket$ (with $\Gamma_{\nu_1,{\bfk_1}} = \Gamma_{\nu_2,{\bfk_2}} = 0$). In this case anharmonic effects enter at second (mixed) order in perturbation theory and can have a significant impact on the scattering rate. 
The transition rate is
\begin{align} \label{eq:TR-2phonon}
  w_{i \to f} &= 2\pi \delta(\omega_{\nu_1,{\bfk_1}}+\omega_{\nu_2,{\bfk_2}}-\omega) \notag \\
  &\times \bigg| \matrixel{p_f;\nu_1,{\bfk_1};\nu_2,{\bfk_2}}{\mathcal{V}(\bfr)}{p_i;0} + \sum\limits_{\nu,\bfk} \bigg( \frac{\matrixel{p_f;\nu,\bfk}{\mathcal{V(\bfr)}}{p_i;0} \matrixel{\nu_1,{\bfk_1};\nu_2,{\bfk_2}}{\delta H}{\nu,\bfk}} {\omega - \omega_{\nu,\bfk} + i\Gamma_{\nu,\bfk}/2} \notag \\
  &+ \frac{\matrixel{\nu,\bfk;\nu_1,{\bfk_1};\nu_2,{\bfk_2}}{\delta H}{0} \matrixel{p_f;0}{\mathcal{V(\bfr)}}{p_i;\nu,\bfk}} {-\omega - \omega_{\nu,\bfk} + i\Gamma_{\nu,\bfk}/2} \bigg) \bigg|^2 \\
  &= 2\pi \delta(\omega_{\nu_1,{\bfk_1}}+\omega_{\nu_2,{\bfk_2}}-\omega) \left(\frac{2\pi b_n}{m_{DM}V}\right)^2 \big|\tilde{F}(q)\big|^2 \notag \\
  &\times\left|\sum_d^\mathfrak{n} A_d e^{-W_d(0)} \left( \mathcal{M}^{(cont)}_{|\nu_1,\bfk_1;\nu_2,\bfk_2\ket,\bfq,d} + \mathcal{M}^{(anh)}_{|\nu_1,\bfk_1;\nu_2,\bfk_2\ket,\bfq,d} \right)\right|^2 \,.
\end{align}
The contact and anharmonic contributions are shown diagrammatically in Fig.~\ref{fig:diagrams}, with the matrix elements given in eqs.~\eqref{eq:contact} \& \eqref{eq:anharmonicM}, and the $\delta H$ matrix element discussed in Sec.~\ref{sec:anharmonic}. 
In the narrow width limit ($\Gamma_{\nu,\bfk}/\omega_{\nu,\bfk}\to0$), and neglecting the interference terms, the anharmonic contribution reduces to the single phonon rate times the branching ratio to $|\nu_1,{\bf k_1};\nu_2,{\bf k_2}\ket$.
Similarly, while eq.~\eqref{eq:TR-2phonon} is strictly only valid for scattering into stable final states, the narrow width approximation applied to multiphonon scattering justifies its use for final states with non-zero width.

\section{Elasticity theory}\label{app:elasticity}
\subsection{The three-phonon Hamiltonian}
In this appendix we briefly review how the leading anharmonic correction to the phonon Hamiltonian can be written in terms of the elasticity parameters, following Refs.~\cite{elasticitybook,Tamura1984}. In elasticity theory, the measure of the size of an infinitesimal deformation of an object is 
\begin{equation}
d\bfx^2 - d\bfa^2 = \left(\frac{\partial x_k}{\partial a_i}da_i\right) \left(\frac{\partial x_k}{\partial a_j}  da_j\right) - da_i da_j = 2 \eta_{ij} da_i da_j \,,
\end{equation}
with $\bfx$ and $\bfa$ the coordinates of a piece of the deformed and undeformed material respectively.  We defined the \emph{Green - St-Venant strain tensor}
\begin{equation}
\eta_{ij} \equiv \frac{1}{2}\left( \frac{\partial x_k}{\partial a_i}\frac{\partial x_k}{\partial a_j}-\delta_{ij} \right) \,,
\end{equation}
which measures how a material responds under stress. Since $x_i=u_i + a_i$ by definition, we can use
\begin{equation}
\frac{\partial x_i}{\partial a_j} = \frac{\partial u_i}{\partial a_j} +\delta_{ij}
\end{equation}
to rewrite the strain tensor as 
\begin{align}
\eta_{ij}&=\frac{1}{2}\left(\frac{\partial u_j}{\partial a_i}+\frac{\partial u_i}{\partial a_j}+\frac{\partial u_k}{\partial a_i}\frac{\partial u_k}{\partial a_j}  \right)\\
&=\frac{1}{2}\left( u_{ij}+u_{ji} + u_{ki} u_{kj}\right) \,, \label{eq:staindefinition}
\end{align}
with $u_{ij}\equiv \partial_i u_j$. Note that $\eta_{ij}$ is manifestly symmetric.

The generalization of Hooke's law is \cite{elasticitybook}
\begin{equation}\label{eq:hooke}
\sigma_{ij} = C_{ijk\ell} \eta_{k\ell} \,,
\end{equation}
with $C_{ijk\ell}$ the elastic constants and $\sigma_{ij}$ the stress tensor. This relation can be written in Hamiltonian form
\begin{equation}\label{eq:hamilsecondorder1}
H = \frac{1}{2}C_{ijk\ell} \eta_{ij}\eta_{k\ell} - \sigma_{ij}\eta_{ij} \,,
\end{equation}
where the stress tensor $\sigma_{ij}$ acts as a source for the $\eta_{ij}$. \eqref{eq:hooke} is then just the equation of motion of $\eta_{ij}$ given by this Hamiltonian. Dropping the source term, the Hamiltonian in \eqref{eq:hamilsecondorder1} can be further generalized to include the cubic response
\begin{equation}\label{eq:hamilthirdorder}
H =\frac{1}{2} C_{ijk\ell} \eta_{ij}\eta_{k\ell} +\frac{1}{3!}C_{ijk\ell m n} \eta_{ij}\eta_{k\ell}\eta_{m n} \,,
\end{equation}
where the $C_{ijk\ell m n}$ are the third order elasticity constants. $C_{ijk\ell}$ is invariant under $i\leftrightarrow j$, $k\leftrightarrow \ell$ and $(ij) \leftrightarrow (k\ell)$, $C_{ijk\ell m n}$ is invariant under $i\leftrightarrow j$, $k\leftrightarrow \ell$, $m\leftrightarrow n$ and the permutations of the $(ij)$, $(k\ell)$ and $(mn)$ pairs. In the most general case, $C_{ijk\ell }$ and $C_{ijk\ell m n}$ have therefore respectively 21 and 56 independent components.

In the isotropic limit, both tensors simplify substantially: $C_{ijk\ell}$ has only 2 independent second order elastic constants, the \emph{Lam\'e parameters} $\mu$ and $\lambda$, which can be related directly to the shear modulus and Young's modulus. The $C_{ijk\ell mn}$ has 3 independent components, parametrized by the third order elastic constants, $\alpha, \beta$ and $\gamma$. Concretely, we can write
\begin{align}
C^{(iso)}_{ijk\ell} &=\lambda \, \delta_{ij}\delta_{k\ell}+\mu \left(\delta_{ik}\delta_{j\ell}+\delta_{i\ell}\delta_{jk}\right) \,, \label{eq:secondorderelastic}\\
C^{(iso)}_{ijk\ell m n}&= \alpha \, \delta_{ij}\delta_{k\ell}\delta_{mn}\nonumber \\
&+ \beta \Big[
\delta_{ij}\left(\delta_{km}\delta_{\ell n}+\delta_{kn}\delta_{\ell m}\right)
+\delta_{k\ell}\left(\delta_{im}\delta_{j n}+\delta_{in}\delta_{j m}\right)
+\delta_{mn}\left(\delta_{ik}\delta_{j\ell}+\delta_{i\ell}\delta_{j k}\right)\Big]\nonumber\\
&+\gamma\Big[ \nonumber
\delta_{nj}\left(\delta_{ik}\delta_{\ell m}+\delta_{i\ell}\delta_{k m}\right)
+\delta_{ni}\left(\delta_{jk}\delta_{\ell m}+\delta_{j\ell}\delta_{k m}\right)
+ \delta_{mj}\left(\delta_{ik}\delta_{\ell n}+\delta_{i\ell}\delta_{k n}\right)\\
&+ \delta_{mi}\left(\delta_{jk}\delta_{\ell n}+\delta_{j\ell}\delta_{k n}\right)\Big] \,,    \label{eq:thirdorderelastic}
\end{align}
where the $\delta_{ij}$ etc.~are Kronecker-$\delta$ symbols. Inserting \eqref{eq:staindefinition}, \eqref{eq:secondorderelastic}  and \eqref{eq:thirdorderelastic} back into \eqref{eq:hamilthirdorder} gives the Hamiltonian in \eqref{eq:tamuraham}. 

\subsection{The isotropic approximation}

The cubic crystals we consider in this work are not completely isotropic but instead are only invariant under permutations of the $x$, $y$ and $z$ axes and parity transformations such as $x\rightarrow -x$ etc. The latter imply that all components of $C^{(cub)}_{ijk\ell}$ and $C^{(cub)}_{ijk\ell m n}$ for which a value of an index occurs an odd number of times must vanish (e.g.~$C^{(cub)}_{1222}=0$ etc). One can show that imposing these symmetries reduces the general elasticity tensors to 3 independent second order elastic constants, and 6 independent third order elastic constants. In order to express the 5 isotropic elasticity parameters $\mu,\lambda, \alpha,\beta$ and $\gamma$ in terms of these 9 measured elasticity parameters for the cubic crystals of interest an averaging procedure is needed.

Given that a 6-tensor such $C^{(cub)}_{ijk\ell m n}$ can be rather unwieldy, much of the literature has chosen to adhere to the \emph{Voigt convention}, where each pair of double indices $(ij)$, $(k\ell)$ and $(mn)$ is replaced with a single index running from 1 to 6 through the mapping
\begin{equation}
\eta_{11}\to\eta_1, \quad\eta_{22}\to\eta_2, \quad\eta_{33}\to\eta_3, \quad\eta_{23}\to\frac{1}{2}\eta_4, \quad\eta_{13}\to\frac{1}{2}\eta_5, \quad\eta_{12}\to\frac{1}{2}\eta_6.
\end{equation}
This maps $C^{(cub)}_{ijk\ell}$ and $C^{(cub)}_{ijk\ell m n}$ to a 2-tensor ($c^{(cub)}_{ij}$) and a 3-tensor ($c^{(cub)}_{ijk}$) respectively, where we have used lowercase $c$ for components of the elasticity tensors in Voigt notation. The independent elasticity parameters for a cubic crystal, as typically reported in the literature, are $c_{11}$, $c_{12}$ and $c_{44}$ for the  second order elastic tensor and $c_{111}$, $c_{112}$, $c_{123}$, $c_{144}$,  $c_{166}$ and $c_{456}$ for the third order elastic tensor\footnote{In certain references $c_{155}$ is reported instead of $c_{166}$; for cubic symmetry $c_{155}=c_{166}$.}, where we have dropped the $(cub)$ superscript going forward. All other components either vanish or can be obtained by applying one of the symmetries listed above. An explicit representation of $c_{ij}$ and $c_{ijk}$ can be found in e.g.~\cite{PhysRevB.76.045202}. 

To obtain the elasticity parameters in the isotropic approximation an averaging procedure must be performed, introducing a certain degree of arbitrariness. We follow the prescription in \cite{Tamura1985}, and define the quantities
\begin{align}
f_2 &= \sum_{i,j,k,\ell} \left(C^{(cub)}_{ijk\ell}- C^{(iso)}_{ijk\ell}\right)^2 \,, \\
f_3 &= \sum_{i,j,k,\ell,m,n} \left(C^{(cub)}_{ijk\ell m n}- C^{(iso)}_{ijk\ell m n}\right)^2 \,,
\end{align}
which provide a measure of the deviation of the isotropic approximation from the cubic case. Minimizing both $f_2$ and $f_3$ leads to the definitions
\begin{align}
\mu&\equiv\frac{1}{5} \left(c_{11} - c_{12} + 3 c_{44}\right) \,,\\
\lambda &\equiv\frac{1}{5} (c_{11} + 4 c_{12} - 2 c_{44}) \,, \\
\alpha&\equiv\frac{1}{35} (c_{111} + 18 c_{112} + 16 c_{123} - 30 c_{144} - 12 c_{166} + 16 c_{456}) \,, \\
\beta&\equiv\frac{1}{35} (c_{111} + 4 c_{112} - 5 c_{123} + 19 c_{144} + 2 c_{166} - 12 c_{456}) \,, \\
\gamma&\equiv\frac{1}{35} (c_{111} - 3 c_{112} + 2 c_{123} - 9 c_{144} + 9 c_{166} + 9 c_{456}) \,,
\end{align}
in agreement with \cite{Tamura1985}.  In the isotropic approximation, the averaged sound speeds of the acoustic phonon modes may also be expressed in terms of $\lambda$, $\mu$, and the mass density $\rho$ as
\begin{align}
    c_{LA} = \sqrt{\frac{\lambda + 2\mu}{\rho}}  \qquad\text{and}\qquad c_{TA} = \sqrt{\frac{\mu}{\rho}} \,.
\end{align}

Both measurements and \emph{ab initio} calculations of the third-order elastic constants are considered rather challenging, and no complete set of experimental results is currently available at close-to-zero temperature. The temperature dependence is mild between room temperature and liquid Nitrogen temperature, but can be large for lower temperatures. For instance, for Ge the combination of $c_{123}+6c_{144}+8c_{456}$ shows a $\mathcal{O}$(100\%) variation between 77K and 3K and even changes sign \cite{Bains1976}. Similarly, the discrepancy between experiment and theory for diamond is also large for $c_{123}$, $c_{144}$ and  $c_{456}$ \cite{elasticdiamond}, presumably due to this temperature dependence. We therefore choose to use the values calculated with Density Functional Theory methods, which are inherently at zero temperature. The values that were used to compute the parameters in Tab.~\ref{tab:elastparam} are listed in Tab.~\ref{tab:cubicelastparam}.

\begin{table}[t]
\begin{ruledtabular}
\begin{tabular}{ccccc}
& Si\footnote{Ref.~\cite{PhysRevB.76.045202} (DFT calculation)\label{foot:DFT1}} & GaAs\footnote{Ref.~\cite{abinitioDFT2} (DFT calculation)\label{foot:DFT2}}& Ge\footnote{Ref.~\cite{elasticGeDFT} (DFT calculation)}  & Diamond\footnote{Ref.~\cite{elasticdiamond} (DFT calculation, measurement)}\\
\hline\hline
$c_{11}$&		153&		126&		129.86&	1051\\
$c_{12}$&		65&		55&		47.39&	125\\
$c_{44}$&		73&		61&		65.73&	560\\\hline
$c_{111}$&	-698&	-600&	-708&	-7611\\
$c_{112}$&	-451&	-401&	-346&	-1637\\
$c_{123}$&	-112&	-94&		-26&		604\\
$c_{144}$&	74&		10&		-10&		-199\\
$c_{166}$&	-253&	-305&	-279&	-2799\\
$c_{456}$&	-57&		-43&		-40&		-1148
\end{tabular}
\end{ruledtabular}
\caption{Elasticity parameters at $T=0K$, in units of GPa. \label{tab:cubicelastparam}}
\end{table}

\allowdisplaybreaks

\section{Exact expressions for long-wavelength structure factors}

\subsection{Anharmonic contributions\label{app:anharmform}}
All expressions below are valid on the domain $0<x<1$, as specified by the Heaviside functions in \eqref{eq:skoAnhLALA}, \eqref{eq:skoAnhTATAout}, \eqref{eq:skoAnhTATAin} and \eqref{eq:skoAnhLATA}. We further defined $\delta\equiv c_{LA}/c_{TA}$. The full expression for the phase space integral for the LA-LA contribution in \eqref{eq:skoAnhLALA} of Sec.~\ref{sec:anharmonic} is then
\begin{align}
g_{LALA}^{(anh)}\left(x\right)&\equiv (2 \beta +4 \gamma +\lambda +3 \mu )^2\frac{\left(x^2-1\right)^3}{2 x^5} \left(x^6+3 x^4+7 x^2+5\right) \left(\tanh ^{-1}(x)-\frac{x^3}{3}-x\right) \nonumber\\
&+ a_{10} x^{10}+a_{8} x^{8}+a_{6} x^{6}+a_{4} x^{4}+a_{2} x^{2}+a_{0} \,,
\intertext{with}
a_{10} &\equiv \frac{1}{6} (2 \beta +4 \gamma +\lambda +3 \mu )^2 \,, \\
a_{8} &\equiv \frac{1}{2} (2 \beta +4 \gamma +\lambda +3 \mu )^2 \,, \\
a_{6} &\equiv -\frac{1}{3} (2 \beta +4 \gamma +\lambda +3 \mu )^2 \,, \\
a_{4} &\equiv \frac{1}{240} \bigg(3 \alpha ^2+2 \alpha  (106 \beta +200 \gamma +53 \lambda +150 \mu )+332 \beta ^2+4 \beta  (120 \gamma +83 \lambda +90 \mu )\nonumber\\
&-320 \gamma ^2+240 \gamma  \lambda -480 \gamma  \mu +83 \lambda ^2+180 \lambda  \mu -180 \mu ^2\bigg) \,, \\
a_{2} &\equiv -\frac{1}{120} \bigg(5 \alpha ^2+2 \alpha  (54 \beta +88 \gamma +27 \lambda +66 \mu )+516 \beta ^2+12 \beta  (136 \gamma +43 \lambda +102 \mu )\nonumber\\
&+1280 \gamma ^2+816 \gamma  \lambda +1920 \gamma  \mu +129 \lambda ^2+612 \lambda  \mu +720 \mu ^2\bigg) \,, \\
a_{0} &\equiv \frac{1}{240} \bigg(15 \alpha ^2+10 \alpha  (10 \beta +8 \gamma +5 \lambda +6 \mu )+668 \beta ^2+4 \beta  (568 \gamma +167 \lambda +426 \mu )\nonumber\\
&+2112 \gamma ^2+1136 \gamma  \lambda +3168 \gamma  \mu +167 \lambda ^2+852 \lambda  \mu +1188 \mu ^2\bigg) \,.
\end{align}
The out-of-plane TA-TA contribution in \eqref{eq:skoAnhTATAout} is given by 
\begin{align}
g_{TATAout}^{(anh)}\left(x\right)&\equiv b_{4} x^{4}+b_{2} x^{2}+b_{0} \,, 
\end{align}
with
\begin{align}
b_{4} &\equiv\frac{43 \beta ^2+2 \beta  (50 \gamma +43 \lambda +50 \mu )+60 \gamma ^2+20 \gamma  (5 \lambda +6 \mu )+43 \lambda ^2+100 \lambda  \mu +60 \mu ^2}{240 } \,, \\
b_{2} &\equiv -\frac{25 \beta ^2+44 \beta  \gamma +50 \beta  \lambda +44 \beta  \mu +20 \gamma ^2+44 \gamma  \lambda +40 \gamma  \mu +25 \lambda ^2+44 \lambda  \mu +20 \mu ^2}{120 } \,, \\
b_{0} &\equiv\frac{15 \beta ^2+10 \beta  (2 \gamma +3 \lambda +2 \mu )+12 \gamma ^2+4 \gamma  (5 \lambda +6 \mu )+15 \lambda ^2+20 \lambda  \mu +12 \mu ^2}{240 }  \,. 
\end{align}
The in-plane TA-TA contribution in \eqref{eq:skoAnhTATAin} is 
\begin{align}
g_{TATAin}^{(anh)}\left(x\right)&\equiv\frac{1}{2}(2 \beta +4 \gamma +\lambda +3 \mu )^2\frac{\left(x^2-1\right)^3 \left(x^2+3\right)}{x} \left(\tanh ^{-1}(x)-\frac{x^3}{3}-x\right)  \nonumber\\
&+ c_{10} x^{10}+c_{8} x^{8}+c_{6} x^{6}+c_{4} x^{4}+c_{2} x^{2}+c_{0} \,, 
\intertext{with}
c_{10} &\equiv \frac{1}{6}(2 \beta +4 \gamma +\lambda +3 \mu )^2 \,, \\
c_{8} &\equiv\frac{1}{2 }(2 \beta +4 \gamma +\lambda +3 \mu )^2 \,, \\
c_{6} &\equiv-\frac{3}{2 } (2 \beta +4 \gamma +\lambda +3 \mu )^2 \,, \\
c_{4} &\equiv\frac{1}{240 }\big(963 \beta ^2+3852 \beta  \gamma +1046 \beta  \lambda +2972 \beta  \mu +3852 \gamma ^2+2092 \gamma  \lambda \nonumber\\
&+5944 \gamma  \mu +283 \lambda ^2+1612 \lambda  \mu +2292 \mu ^2\big) \,, \\
c_{2} &\equiv-\frac{1}{24 }\big(17 \beta ^2+68 \beta  \gamma +26 \beta  \lambda +60 \beta  \mu +68 \gamma ^2+52 \gamma  \lambda\nonumber\\
& +120 \gamma  \mu +9 \lambda ^2+44 \lambda  \mu +52 \mu ^2\big) \,, \\
c_{0} &\equiv\frac{1}{16}(\beta +2 \gamma +\lambda +2 \mu )^2 \,.  
\end{align}
Finally, the LA-TA contribution is given by the piecewise function
\begin{align}
g_{LATA}^{(anh)}(x)&\equiv\left\{\begin{array}{ll}
g_{LATA,1}^{(anh)}(x) &\text{if}\; 0<x<\frac{1}{\delta} \,,\\
g_{LATA,2}^{(anh)}(x) &\text{if}\; \frac{1}{\delta}<x<1 \,,
\end{array}\right.
\intertext{where}
g_{LATA,1}^{(anh)}(x) &\equiv\frac{(2 \beta +4 \gamma +\lambda +3 \mu )^2}{2  (\delta +1)^5}\Bigg[
-(\delta +1)^5 \frac{\left(x^2-1\right)^3 \left(x^2+3\right)}{x} \left(\tanh ^{-1}(x)-\frac{x^3}{3}-x\right)\nonumber\\
&-\frac{(\delta +1)^5}{\delta^{12}} \frac{\left(\delta ^2 x^2-1\right)^3 \left(\delta ^6 x^6+3 \delta ^4 x^4+7 \delta ^2 x^2+5\right)}{x^5} \left(\tanh ^{-1}(\delta  x)-\frac{1}{3} \delta ^3 x^3-\delta  x\right)\nonumber\\
&+ d_{10} x^{10}+d_{8} x^{8}+d_{6} x^{6}+d_{4} x^{4}+d_{2} x^{2}+d_{0}\Bigg] \,, 
\intertext{with}
d_{10} &\equiv-\frac{1}{3} (\delta +1)^6 \left(\delta ^2-\delta +1\right) \,, \\
d_{8} &\equiv -(\delta +1)^6 \,, \\
d_{6} &\equiv\frac{1}{3 \delta }(\delta +1)^5 (9 \delta +2) \,, \\
d_{4} &\equiv-\frac{1}{315 \delta ^3}\big(189 \delta ^8+945 \delta ^7+2706 \delta ^6+5340 \delta ^5+5779 \delta ^4\nonumber\\&
+1505 \delta ^3-2460 \delta ^2-1870 \delta -374\big) \,, \\
d_{2} &\equiv\frac{1}{105 \delta ^5}\big(-32 \delta ^6+365 \delta ^5+1057 \delta ^4+930 \delta ^3+930 \delta ^2+465 \delta +93\big) \,, \\
d_{0} &\equiv-\frac{1}{15 \delta ^7}\big(-16 \delta ^6+15 \delta ^5+75 \delta ^4+150 \delta ^3+150 \delta ^2+75 \delta +15\big) \,,
\end{align}
and
\begin{align}
g_{LATA,2}^{(anh)}(x) &\equiv\frac{(2 \beta +4 \gamma +\lambda +3 \mu )^2}{2 \delta ^{12} \left(\delta ^2-1\right)^5 x^5}\Bigg[-\left(\delta ^2-1\right)^5 \left(\delta ^2x^2-1\right)^3 \left(\delta ^6 x^6+3 \delta ^4 x^4+7 \delta ^2 x^2+5\right)\nonumber \\
&\times\coth ^{-1}(\delta  x)+\left(\delta ^2-1\right)^5  \big[\left(6 \delta ^{12}+\delta ^8\right) x^8-8 (\delta ^{12}+\delta ^6) x^6+3 \left(\delta ^{12}+\delta ^4\right) x^4\nonumber\\&
+8 \delta ^2 x^2-5\big]\coth ^{-1}(\delta )+ \sum_{i=1}^{11} e_i x^i\Bigg] \,, 
\intertext{with}
e_{11} &\equiv\delta ^{11} \left(\delta ^2-1\right)^5 \,, \\
e_{10} &\equiv0  \,, \\
e_{9} &\equiv\frac{ \delta ^9}{315} \left(105 \delta ^{10}-861 \delta ^8+3066 \delta ^6-4266 \delta ^4+525 \delta ^2+151\right) \,, \\
e_{8} &\equiv\frac{\delta ^9}{3}  \left(-18 \delta ^{12}+84 \delta ^{10}-147 \delta ^8+74 \delta ^6+82 \delta ^4-14 \delta ^2+3\right) \,, \\
e_{7} &\equiv-\frac{2\delta ^7}{105}  \left(105 \delta ^{10}-1645 \delta ^8+5474 \delta ^6-2914 \delta ^4+1605 \delta ^2-321\right) \,, \\
e_{6} &\equiv\frac{8\delta ^7}{3}  \left(3 \delta ^{14}-14 \delta ^{12}+26 \delta ^{10}-29 \delta ^8+43 \delta ^6-24 \delta ^4+14 \delta ^2-3\right) \,, \\
e_{5} &\equiv-\frac{2\delta ^5}{15}  \left(5 \delta ^{10}+255 \delta ^8-342 \delta ^6+350 \delta ^4-175 \delta ^2+35\right) \,, \\
e_{4} &\equiv\frac{\delta ^5 }{15} \big(-45 \delta ^{16}+210 \delta ^{14}-384 \delta ^{12}+334 \delta ^{10}+16 \delta ^8-350 \delta ^6\nonumber\\
&+384 \delta ^4-210\delta ^2+45\big) \,, \\
e_{3} &\equiv\frac{19\delta ^3}{3}  \left(\delta ^2-1\right)^5 \,, \\
e_{2} &\equiv-\frac{8\delta ^3}{105}  \left(34 \delta ^{10}-329 \delta ^8+790 \delta ^6-896 \delta ^4+490 \delta ^2-105\right) \,, \\
e_{1} &\equiv-5 \delta  \left(\delta ^2-1\right)^5 \,, \\
e_{0} &\equiv\frac{64 \delta ^{11}}{35}-\frac{965 \delta ^9}{63}+\frac{790 \delta ^7}{21}-\frac{128 \delta ^5}{3}+\frac{70 \delta ^3}{3}-5 \delta \,.
\end{align}

From the matrix element in Eq.~\eqref{eq:tamuramatrixelem}, the widths for each anharmonic channel may be calculated explicitly, giving 
\begin{align}
 \Gamma_{\text{LA}\to\text{LALA}}\left( q \right) &= \frac{q^5}{960 \pi {c_{LA}}^4 \rho^3} \left(\alpha + 6\beta + 8\gamma + 3\lambda + 6\mu \right)^2 \,, \intertext{and}
\Gamma_{\text{LA}\to\text{TATAin}} \left(q \right) &= \frac{q^5}{7680\pi {c_{LA}}^4 \rho^3}\left(f_{1}\left(\delta^2 -1\right)^3 \left(1 + 3\delta^2\right) \coth^{-1}\left(\delta\right) + \sum_{i=1}^{i=4} f_i \delta^{2i-1} \right) \,, \intertext{with}
f_{4} &\equiv 15(97 \beta^2 + 388 \beta \gamma + 388 \gamma^2 + 98 \beta \gamma + 196 \gamma \lambda + 25\lambda^2 \nonumber\\
&+ 4 \mu \left(73 \left(\beta + 2\gamma \right) + 37\lambda \right) + 220 \mu^2 ) \,, \\
f_{3} &\equiv -10 ( 353 \beta^2 + 1412 \gamma^2 + 724 \gamma \lambda + 93 \lambda^2 + 2136 \gamma \mu + 548 \lambda \mu \nonumber\\
& + 808 \mu^2 + 2\beta \left( 706 \gamma + 181 \lambda + 534 \mu \right) ) \,, \\
f_{2} &\equiv  2563\beta^2 + 10252 \gamma^2 + 5292 \gamma\lambda + 683 \lambda^2 + 15544 \gamma \mu + 4012 \lambda\mu\nonumber\\
&+ 5892\mu^2 + 2\beta\left( 5126 \gamma + 1323 \lambda + 3886\mu \right) \,, \\
f_{1} &\equiv  -120\left(2\beta +4\gamma + \lambda + 3\mu \right)^2 \,, \intertext{and}
\Gamma_{\text{LA}\to\text{TATAout}} \left(q \right) &= \frac{q^5}{7680 \pi {c_{{LA}}}^4 \rho^3}  \sum_{i=1}^{i=4} g_i \delta^{2i-1} \,, 
\intertext{with}
g_{4}&\equiv 15\beta^2 + 20\beta\gamma + 12\gamma^2 + 30\beta\lambda + 20\gamma\lambda + 15\lambda^2 \nonumber\\
&+4\mu \left(5\beta + 6\gamma + 5\lambda \right) + 12\mu^2 \,, \\
g_{3} &\equiv -2 (25 \beta^2 + 20\gamma^2 + 44\gamma\lambda + 25\lambda^2 + 40 \gamma\mu + 44\lambda\mu \nonumber\\
&+ 20\mu^2 + \beta \left( 44 \gamma + 50\lambda + 44\mu \right)) \,, \\
g_{2} &\equiv 43\beta^2 + 100\beta \gamma + 60\gamma^2 + 86 \beta \lambda + 100\gamma \lambda + 43\lambda^2 \nonumber\\
&+20\mu \left(5 \beta +6\gamma + 5\lambda \right) + 60\mu^2 \,, \\
g_{1}&\equiv0 \,, 
\end{align}
and finally
\begin{align}
\Gamma_{\text{LA}\to\text{LATA}} \left(q \right) &= \frac{ h_{12} q^5}{64\pi {c_{LA}}^4 \rho^3}\left( \left(\delta^2 - 1\right)^3 \left(1 + 3\delta^2 \right) \coth^{-1} \left(\delta \right) + \frac{\left(1-\delta\right)}{315 \left( 1+ \delta \right)^5} \sum_{i=0}^{i=11} h_i \delta^i \right) \,, \intertext{with}
h_{12}&\equiv \left(2 \beta + 4 \gamma + \lambda + 3\mu \right)^2 \,, \\
h_{11}&\equiv 945 \,, \\
h_{10}&\equiv 5670 \,, \\
h_{9}&\equiv 12915 \,, \\
h_{8}&\equiv 11340 \,, \\
h_{7}&\equiv -4746 \,, \\
h_{6}&\equiv -19656 \,, \\
h_{5}&\equiv -18030 \,, \\
h_{4}&\equiv -6540 \,, \\
h_{3}&\equiv 793 \,, \\
h_{2}&\equiv 2658 \,, \\
h_{1}&\equiv 1083 \,, \\
h_{0}&\equiv 128 \,. 
\end{align}

\subsection{Contact contributions\label{app:contact}}
The functions parametrizing the phase space integrals in Sec.~\ref{sec:contactterm} can be expressed as 
\begin{align}
g_{LALA}^{(cont)}\left(x\right)&\equiv \frac{-x(x^6+x^4-x^2-3)+(x^8+2x^4-3)\tanh^{-1}(x)}{x^5} \,, \\
g_{TATA}^{(cont)}\left(x\right)&\equiv\frac{(1-x^2)^2}{x^5}\left(x(3-x^2)+(x^4+2x^2-3)\tanh^{-1}(x)\right) \,. \\
\intertext{The LA-TA mode is given by the piecewise function}
g_{LATA}^{(cont)}(x)&\equiv\left\{\begin{array}{ll}
g_{LATA,1}^{(cont)}(x) &\text{if}\; 0<x<\frac{1}{\delta} \,, \\
g_{LATA,2}^{(cont)}(x) &\text{if}\; \frac{1}{\delta}<x<1 \,,
\end{array}\right.
\intertext{with}
g_{LATA,1}^{(cont)}\left(x\right)&\equiv-\frac{ \delta  (\delta +1)}{ x^5}\bigg[ \left(x^2+3\right) \left(x^2-1\right)^3 \tanh ^{-1}(x)+\left(x^8+ \frac{2x^4}{\delta^4}-\frac{3}{\delta^{8}}\right) \tanh ^{-1}(\delta  x)\bigg]\nonumber\\
&+\frac{1}{15 \delta^6 x^4}\bigg[15 (\delta +1)^2 \delta^6 x^6+\left(-59 \delta^4-59 \delta^3+16 \delta^2+21 \delta +21\right) \delta^4 x^4\nonumber\\
&+15 \left(7 \delta^6+7 \delta^5-\delta -1\right) \delta^2 x^2-45 \left(\delta^8+\delta^7+\delta +1\right)\bigg] \,, \\
g_{LATA,2}^{(cont)}\left(x\right)&\equiv-\frac{ \delta(\delta +1)}{ x^5}\bigg[ \left(8 x^2-\left(\frac{2}{\delta^4}+6\right) x^4-3+\frac{3}{\delta^8}\right)\coth ^{-1}(\delta )\nonumber\\&
+\left(x^8+\frac{2 x^4}{\delta^4}-\frac{3}{\delta^8}\right) \coth ^{-1}(\delta  x)\bigg]+\frac{1}{15\delta^7(\delta-1)x^5}\bigg[15 \delta^7 \left(\delta^2-1\right) x^7\nonumber\\
&+ \delta^5 \left(5 \delta^2-21\right) x^5+30 \delta^5\left(-3 \delta^4+2 \delta^2+1\right)  x^4-15 \delta^3 \left(\delta^2-1\right) x^3\nonumber\\
&+40 \delta^7\left(3 \delta^2-2\right)  x^2-45  \delta\left(\delta^2-1\right)  x-45 \delta^9+30 \delta^7+6 \delta^5+30 \delta^3-45 \delta
\bigg] \,.
\end{align}
All functions are only to be evaluated for $0<x<1$, as enforced by the Heaviside functions in \eqref{skoContLALA}, \eqref{skoContTATA} and \eqref{skoContLATA}.

\bibliographystyle{apsrev4-1}
\bibliography{phonons}

\end{document}